\documentclass[pra,preprint,aps,eqsecnum,showpacs]{revtex4-1}
\usepackage{graphics}
\usepackage{amsmath}
\usepackage{amssymb}
\newtheorem{theorem}{Theorem}
\newtheorem{lemma}{Lemma}
\begin{document}
\title[]
{Invariant theoretic approach to uncertainty relations for quantum systems}
 \author{J. Solomon Ivan} 
\email{solomon@rri.res.in} 
\affiliation{Raman Research Institute, C. V. Raman Avenue,
  Sadashivanagar, Bangalore 560 080.}  
\author{Krishna Kumar Sabapathy} 
\email{kkumar@imsc.res.in} 
\affiliation{The Institute of 
 Mathematical Sciences, C. I. T. Campus, Chennai 600113.} 
\author{N. Mukunda} 
\email{nmukunda@gmail.com} 
\affiliation{The Institute of 
 Mathematical Sciences, C. I. T. Campus, Chennai 600113.} 
\author{R. Simon} 
\email{simon@imsc.res.in} 
\affiliation{The Institute of 
 Mathematical Sciences, C. I. T. Campus, Chennai 600113.} 
\begin{abstract} 
We present a general framework and procedure to derive uncertainty
relations for observables of quantum systems in a covariant
manner. All such relations are consequences of the positive
semidefiniteness of the density matrix of a general quantum
state. Particular emphasis is given to the action of unitary symmetry
operations of the system on the chosen observables, and the covariance
of the uncertainty relations under these operations. The general
method is applied to the case of an $n$-mode system to recover the
 $Sp(2n,\,R)$-covariant multi mode generalization of the single mode 
Schr\"{o}dinger-Robertson Uncertainty Principle; and to the set of all 
polynomials in canonical
variables for a single mode system. In the latter situation, the case
of the fourth order moments is analyzed in detail, exploiting
covariance under the homogeneous Lorentz group $SO(2,\,1)$ of which the symplectic 
group $Sp(2,\,R)$ is the double cover.    
\end{abstract}
\maketitle 
 
\section{Introduction}
It is a well known historical fact that the 1925\,--\,1926
discoveries of two equivalent mathematical formulations of
quantum mechanics---Heisenberg's matrix form followed by
Schr\"{o}dinger's wave mechanical form---preceded the development of a
physical interpretation of these formalisms\,\cite{sch-hei}. 
The first important
ingredient of the conventional interpretation was Born's 1926
identification of the squared modulus of a complex Schr\"{o}dinger
wavefunction as a probability\,\cite{born26}. The second ingredient developed in 1927
was Heisenberg's Uncertainty Principle (UP)\,\cite{heisenbergup}. 
To these may be added
Bohr's Complementarity Principle which has a more philosophical
flavour\,\cite{bohrcp}.

Heisenberg's original derivation of his position-momentum UP combined
the formula for the resolving power of an optical microscope
extrapolated to a hypothetical gamma ray microscope, with the energy
and momentum relations for a single photon, in analysing the inherent
limitations in simultaneous determinations of the position and
momentum of an electron. His result indicated the limits of
applicability of classical notions, in particular the spatial orbit of
a point particle, in quantum mechanics.

More formal mathematical derivations of the UP, using the Born
probability interpretation, soon followed. Prominent among them are
the treatments of Kennard, Schr\"{o}dinger, and Robertson\,\cite{uncer0}. Such a
derivation was also presented by Heisenberg in his 1930 Chicago 
lectures\,\cite{heisenberg-chicago}.  

The Heisenberg position-momentum UP is basically kinematical in
nature. In contrast, the Bohr UP for time and energy involves quantum
dynamics in an essential manner\,\cite{bohr-te-up}. Later work on the UP has introduced a
wide variety of ideas\,\cite{ideas} and interpretations of the fluctuations or the
uncertainties involved\,\cite{interpretUP}, such as in entropic\,\cite{EUP} and other formulations\,\cite{otherUP}.

Even for a one-dimensional quantum system, the Schr\"{o}dinger-Robertson form of 
the
UP displays more invariance than the Heisenberg form. Thus while the
 latter is invariant only under reciprocal scalings of position and
momentum, and their interchange amounting to Fourier transformation, the
former is invariant under the three-parameter Lie group $Sp(2,\,R)$
of linear canonical transformations. Fourier transformation, as well
as reciprocal scalings, belong to $Sp(2,\,R)$\,\cite{simon88}. The generalisation of
the Schr\"{o}dinger-Robertson UP to any finite number, $n$, of degrees of freedom
displays invariance under the group $Sp(2n,\,R)$\,\cite{dutta94}. 

The purpose of this
paper is to outline an invariant theoretic approach to general
uncertainty relations for quantum systems. It combines a recapitulation
and reexpression of some past results\,\cite{moments2012} with some new ones geared to
practical applications. The analysis throughout is in the spirit 
of the Schr\"{o}dinger-Robertson
treatment, and, in particular, our considerations do not cover the 
entropic type uncertainty relations. All our considerations will be kinematical in
nature.

The material of this paper is presented as follows. Section II sets up
a general framework and procedure for deriving consequences of the
positive semidefiniteness of the density matrix of a general quantum
state, for the expectation values and fluctuations of a chosen
(linearly independent) set of observables for the system. This has the
form of a general uncertainty relation. A natural way to separate the 
expressions entering it into  a symmetric fluctuation part, and an
antisymmetric part contributed by commutators among the observables,
hence specifically quantum in origin, is described. With respect to any
unitary symmetry operation associated with the system, under which the
chosen observables transform in a suitable manner, the uncertainty
relation is shown to transform covariantly and to be preserved in
content. In Section III this general framework is applied to the case
of a quantum system involving $n$ Cartesian canonical Heisenberg
pairs, i.e., an $n$-mode system\,; and to the fluctuations in
canonical `coordinates' and `momenta' in any state. The resulting
 $n$-mode generalization of the original
Schr\"{o}dinger-Robertson UP is seen to be explicitly covariant under
the group $Sp(2n,\,R)$ of linear homogeneous canonical
transformations. Section IV returns to the single mode system, but
considers as the system of observables the infinite set of operator polynomials of all
orders in the two canonical variables. The treatment is formal to the
extent that unbounded operators are involved. An important role is
played by the set of all finite-dimensional real nonunitary
irreducible representations of the covariance group $Sp(2,\,R)$. We
follow in spirit the structure of the basic theorems in the classical
theory of moments. Thus the formal infinite-dimensional matrix
uncertainty relation is reduced to a nested sequence of finite-dimensional requirements, of steadily increasing dimensions. 
While this case has been treated elsewhere\,\cite{moments2012}, some of the subtler
aspects are now carefully brought out. 
 In this and the subsequent Sections the method of Wigner distributions is used 
 as an extremely convenient technical tool.  
Section V treats in more detail the
uncertainty relations of Section IV that go one step beyond  the
original Schr\"{o}dinger-Robertson UP. Here all the fourth order moments
of the canonical variables in a general state are involved. Their
fully covariant treatment brings in the defining and some other low
dimensional representations of the three-dimensional Lorentz group
$SO(2,\,1)$. It is shown that the uncertainty relations (to the
concerned order) are all expressible in terms of $SO(2,\,1)$
invariants. 
 In Section VI we describe an interesting aspect of the Schr\"{o}dinger-Robertson 
 UP in the light of three-dimensional Lorentz geometry, which becomes particularly 
 apparent through the use of Wigner distribution methods. We argue that this 
should generalise to the conditions on fourth (and higher) order moments as well. 
The paper ends with some concluding remarks in Section VII.

\section{General Framework}

We consider a quantum system with associated Hilbert space ${\cal H}$,
state vectors $|\psi \rangle$, $|\phi \rangle$, $\cdots$ and inner
product $\langle \phi | \psi \rangle$ as usual. A general (mixed)
state is determined by a density operator or density matrix
$\hat{\rho}$ acting on ${\cal H}$ and obeying 
\begin{eqnarray}
\hat{\rho}^{\dagger} =\hat{\rho} \geq 0,\,\,\,\,\,{\rm Tr}\,\hat{\rho}=1.
\label{2.1}
\end{eqnarray}    
Then ${\rm Tr}\,\hat{\rho}^2=1$ or $< 1$ distinguishes between pure and
mixed states. Any hermitian observable $\hat{A}$ of the system
possesses the expectation value
\begin{eqnarray}
\langle \hat{A} \rangle = {\rm Tr}\,(\hat{\rho}\,\hat{A})
\label{2.2}
\end{eqnarray}
in the state $\hat{\rho}$, the dependence of the left hand side on
$\hat{\rho}$ being generally left implicit.

We now set up a general method which allows the drawing out of the
consequences of the nonnegativity of $\hat{\rho}$ in a systematic
manner. This along with two elementary lemmas will be the basis of our
considerations. 

Let $\hat{A}_{a}$, $a=1,\,2,\,\cdots,\,N$ be a set of $N$ {\em linearly}
independent {\em hermitian} operators, each representing some
observable of the system. We set up two formal $N$-component and
$(N+1)$-component column vectors with hermitian operator entries as
follows\,:
\begin{eqnarray}
\hat{A}=\left( 
\begin{array}{c}
\hat{A}_{1} \\
\vdots \\
\vdots \\
\hat{A}_{N}
\end{array}
\right),
\,\,\,\,\,
\hat{{\cal A}}=\left( \begin{array}{c} 1 \\ \hat{A} \end{array}\right)
= \left( 
\begin{array}{c}
1\\
\hat{A}_{1} \\
\vdots \\
\vdots \\
\hat{A}_{N}
\end{array}
\right),
\label{2.3}
\end{eqnarray} 
From $\hat{{\cal A}}$ we construct a square $(N+1)$-dimensional `matrix' with 
operator entries as 
\begin{eqnarray}
\hat{\Omega} = \hat{{\cal A}}\hat{{\cal A}}^{T} =
\left(\begin{array}{ccccc}
1 & \cdots & \cdots & \hat{A}_{b} & \cdots \\
\vdots &&&\vdots&\\
\hat{A}_{a}&\cdots & \cdots & \hat{A}_{a}\hat{A}_{b} & \cdots \\
\vdots &&&\vdots&
\end{array}
\right).
\label{2.4}
\end{eqnarray}
Since $(\hat{A}_{a}\hat{A}_{b} )^{\dagger}= \hat{A}_{b}\hat{A}_{a}$,
$\hat{\Omega}$ is `hermitian' in the following sense\,: taking the
operator hermitian conjugate of each element and then transposing the
rows and columns leaves $\hat{\Omega}$ unchanged. In a state
$\hat{\rho}$ we then have an $(N+1)$-dimensional numerical hermitian
matrix $\Omega$ of the expectation values of the elements of
$\hat{\Omega}$\,:
\begin{eqnarray}
\Omega = \langle \hat{\Omega} \rangle &=& {\rm
  Tr}(\hat{\rho}\,\hat{\Omega})=
\left(\begin{array}{ccccc}
1 & \cdots & \cdots & \langle \hat{A}_{b}\rangle & \cdots \\
\vdots &&&\vdots&\\
\langle \hat{A}_{a}\rangle &\cdots & \cdots & 
       \langle \hat{A}_{a}\hat{A}_{b}\rangle & \cdots \\
\vdots &&&\vdots&
\end{array}
\right), \nonumber \\
{\rm i.e.}, ~ \Omega_{ab} &=& {\rm Tr}(\hat{\rho}\,\hat{\Omega}_{ab})\,;\nonumber\\
\Omega^{\dagger}&=& \Omega.
\label{2.5}
\end{eqnarray}
Now for any complex $(N+1)$ component column vector ${\bf C}=
(c_0,\,c_1,\, \cdots, \, c_N)^T$ we have
\begin{eqnarray}
{\bf C}^{\dagger}\,\hat{\Omega}\,{\bf C}&=& 
{\bf C}^{\dagger}\hat{{\cal 
A}}\,({\bf C}^{\dagger}\hat{{\cal A}})^{\dagger}\geq 0, \nonumber \\
\langle {\bf C}^{\dagger}\,\hat{\Omega}\,{\bf C} \rangle
&=& {\bf C}^{\dagger}\,{\Omega}\,{\bf C} \geq 0,
\label{2.6}
\end{eqnarray}
leading immediately to\,:
\begin{theorem} Positivity of $\hat{\rho}$ imputes positivity 
to the matrix $\Omega$, for every choice of $\hat{{\cal A}}$\,:
{\begin{equation}
\hat{\rho}\geq 0 \Rightarrow \Omega = \langle \hat{\Omega} \rangle = {\rm
  Tr}(\hat{\rho}\,\hat{\Omega})\geq 0\,, ~~ \forall \,\hat{\cal A}\,. 
\label{2.7}
\end{equation}}
\end{theorem}
This is thus an uncertainty relation valid in every physical state
$\hat{\rho}$.

\vskip0.2cm
\noindent
{\bf Remark}\,: It is for the sake of definiteness and keeping in view the ensuing
applications that we have assumed the entries $\hat{{A}}_{a}$ of 
$\hat{\cal A}$ and $\hat{A}$ to be all hermitian. This can be relaxed
and each $\hat{A}_{a}$ can be a general linear operator pertinent to the
system. The only change would be the replacement of $\hat{\cal A}^{T}$
in Eq.\,(\ref{2.4}) by $\hat{\cal A}^{\dagger}$, leading to a result
similar to Theorem 1.

\vskip0.2cm

Depending on the basic kinematics of the system we can imagine various
choices of the $\hat{A}_{a}$ geared to exhibiting corresponding
symmetries or covariance properties of the uncertainty relation
(\ref{2.7}). Specifically suppose there is a unitary operator $\overline{U}$
on ${\cal H}$ such that under  conjugation the $\hat{A}_{a}$ go into 
 (necessarily real) linear combinations of themselves\,:
\begin{eqnarray} 
\overline{U}\,\overline{U}^{\,\dagger}&=& 
\overline{U}^{\,\dagger}\,\overline{U}=1\!\!1,
\nonumber \\
\overline{U}^{\,-1}\,\hat{A}_{a}\,\overline{U}&=& R_{ab}\,\hat{A}_{b}, \nonumber
\\
\overline{U}^{\,-1}\,\hat{\cal A}\,\overline{U}&=& {\cal R} \,\hat{\cal A},
\nonumber \\
{\cal R}&=& \left( 
\begin{array}{cc}
1 & 0 \\
0 & R
\end{array}
\right), \,\,\,\, R=(R_{ab}).
\label{2.8}
\end{eqnarray}
The matrix $R$ here is real $N$-dimensional nonsingular. Then combined
with Eq.\,(\ref{2.5}) we have\,:
\begin{eqnarray}
\hat{\rho}^{\, \prime} = \overline{U}\,\hat{\rho}\,\overline{U}^{\,-1} \Rightarrow \Omega^{\, \prime}
&=&{\rm Tr}(\hat{\rho}^{\, \prime}\,\hat{\Omega})={\rm
  Tr}(\hat{\rho}\,\overline{U}^{\,-1}\,\hat{\cal A}\,\hat{\cal A}^{T}\,\overline{U})
\nonumber \\
&=& {\cal R} \,\Omega \,{\cal R}^{T}, \nonumber \\
\Omega \geq 0 &\Leftrightarrow& \Omega^{\, \prime} \geq 0.
\label{2.9}
\end{eqnarray}
This is because the passage  $\Omega \rightarrow \Omega^{\, \prime}$ is a
congruence transformation. Thus the uncertainty relation (\ref{2.7}) is
covariant or explicitly preserved under the conjugation of the state
$\hat{\rho}$ by the unitary transformation $\overline{U}$.

We now introduce two lemmas concerning (finite-dimensional) nonnegative
matrices, whose proofs are elementary\,:
\begin{lemma}
For a hermitian positive definite matrix in block form,
\begin{eqnarray}
Q=Q^{\dagger}= \left( \begin{array}{cc}
A & C^{\dagger} \\
C & B
\end{array} \right),\,\,\,A^{\dagger}=A\,,\;\,\,B^{\dagger}= B,
\label{2.10}
\end{eqnarray}
we have 
\begin{eqnarray}
Q > 0 \,\, \Leftrightarrow \,\, A > 0\,\,\,{\rm
  and}\,\,\,B-C\,A^{-1}C^{\dagger} >0.
\label{2.11}
\end{eqnarray}
\end{lemma}
The proof consists in noting that by a congruence we can pass from $Q$
to a block diagonal form\,\cite{hjbook}\,:
\begin{eqnarray}
Q = \left( \begin{array}{cc}
1\!\!1 & 0 \\
-CA^{-1} & 1\!\!1
\end{array} \right)
\left( \begin{array}{cc}
A & 0 \\
0 & B- C A^{-1}C^{\dagger}
\end{array} \right)
\left( \begin{array}{cc}
1\!\!1 & 0 \\
-CA^{-1} & 1\!\!1
\end{array} \right)^{\dagger}.
\label{2.12}
\end{eqnarray}
\begin{lemma}
If we separate a hermitian matrix $Q$ into real symmetric and pure
imaginary antisymmetric parts $R,\,S$  then
\begin{eqnarray}
Q=Q^{\dagger} = R + iS\geq 0, \,\,{\rm det}\,S \not= 0 \,\Rightarrow R
> 0.
\label{2.13}
\end{eqnarray}
\end{lemma}
The nonsingularity of $S$ means that $Q$ must be even
dimensional. (The proof, which is elementary, is omitted).

Now we apply Lemma 1 to the $(N+1)$-dimensional matrix $\Omega$ in
Eq.\,(\ref{2.5}), choosing a partitioning where $B$ is $N \times N$,
$C$ is $N \times 1$ and $C^{\dagger}$ is $1 \times N$\,:
\begin{eqnarray}
\Omega = \left( \begin{array}{cc}
A & C^{\dagger} \\
C & B
\end{array} \right):\,\, A=1,\,\,B=(\langle
\hat{A}_{a}\hat{A}_{b}\rangle),\,\,\,C=(\langle \hat{A_a} \rangle).
\label{2.14}
\end{eqnarray}  
Then from Eq.\,(\ref{2.11}) we conclude\,:
\begin{theorem}
\begin{eqnarray}
\hat{\rho}\geq 0 &\Rightarrow & \Omega \geq 0 \Leftrightarrow
\nonumber \\
\tilde{\Omega}&=& (\langle(\hat{A}_{a} - \langle\hat{A}_{a}
\rangle)(\hat{A}_{b}- \langle \hat{A}_{b} \rangle) \rangle ) \geq 0.
\label{2.15}
\end{eqnarray}
\end{theorem}
All expectation values involved in the elements of 
the $N\times N$ matrix $\tilde{\Omega}$ are
with respect to the state $\hat{\rho}$.

The motivation for the definitions of $\hat{\cal A}$, $\hat{\Omega}$
as in Eq.\,(\ref{2.3}) is now clear\,: after an application of Lemma 1 we
immediately descend from the matrix $\Omega$ to the matrix
$\tilde{\Omega}$ involving only expectation values of products of
deviations from means. It is then natural to write the elements
of $\tilde{\Omega}$ as follows\,:
\begin{eqnarray}
\Delta \hat{A}_{a}&=& \hat{A}_{a}- \langle \hat{A}_{a} \rangle,
\nonumber \\
\tilde{\Omega}_{ab}&=& \langle\Delta \hat{A}_{a}\Delta \hat{A}_{b} \rangle.
\label{2.16}
\end{eqnarray}
We revert to this form shortly.

The covariance of the statement (\ref{2.15}), Theorem 2, under a unitary
symmetry $\overline{U}$ acting as in Eq.\,(\ref{2.8}) follows from a brief
calculation\,: 
\begin{eqnarray}
\hat{\rho} \rightarrow \hat{\rho}^{\, \prime} = \overline{U} \hat{\rho}\,\overline{U}^{\,-1} 
 & \Rightarrow & \nonumber \\
\overline{U}^{\,-1} (\hat{A}_{a} - {\rm Tr}(\hat{\rho}^{\, \prime} \hat{A}_{a}))\overline{U}
&=&\overline{U}^{\,-1} \hat{A}_{a} \overline{U} -{\rm
  Tr}(\hat{\rho}\,\overline{U}^{\,-1}\hat{A}_{a} \overline{U}) \nonumber \\
&=& R_{ab}(\hat{A}_{b}- {\rm Tr}(\hat{\rho}\hat{A}_{b})); \nonumber \\
\tilde{\Omega}\rightarrow \tilde{\Omega}^{\, \prime}&=& R \tilde{\Omega} R^{T};
\nonumber \\
\tilde{\Omega} \geq 0 & \Leftrightarrow & \tilde{\Omega}^{\, \prime} \geq 0.
\label{2.17}
\end{eqnarray}

We now return to Eq.\,(\ref{2.16}). The state $\hat{\rho}$ being kept
fixed, we can split the hermitian $N \times N$ matrix $\tilde{\Omega}$
into real symmetric and pure imaginary antisymmetric parts as follows\,:
\begin{eqnarray}
\tilde{\Omega}_{ab} &=& V_{ab}(\hat{\rho}\,;\,\hat{A})+
\frac{i}{2}\,\omega_{ab}(\hat{\rho}\,; \,\hat{A}), \nonumber \\
V_{ab}(\hat{\rho}\,;\,\hat{A})&=&V_{ba}(\hat{\rho}\,;\,\hat{A})=
\frac{1}{2}\langle\{\Delta \hat{A}_{a}, \, \Delta \hat{A}_{b} \}
\rangle \nonumber \\
&=& \frac{1}{2}\langle\{ \hat{A}_{a}, \,  \hat{A}_{b} \}
\rangle - \langle \hat{A}_{a}\rangle \langle  \hat{A}_{b} \rangle;
\nonumber \\
\omega_{ab}(\hat{\rho}\,;\,\hat{A})&=&- \omega_{ba}(\hat{\rho}\,;\,\hat{A})=
 \,- \, i\, \langle [ \hat{A}_{a}, \,  \hat{A}_{b}] \rangle .
\label{2.18}
\end{eqnarray}
The brackets $[\,\cdot\,,\,\cdot\,]$ and $\{\,\cdot\,,\,\cdot\,\}$ denote, as usual, the commutator 
and anticommutator respectively. 
The natural physical identification of the $N \times N$ real symmetric
matrix $V(\hat{\rho}\,;\,\hat{A})=(V_{ab}(\hat{\rho}\,;\,\hat{A}))$ is
that it is the variance matrix (or matrix of covariances) associated 
with the set $\{\hat{A}_{a}
\}$ in the state $\hat{\rho}$. The uncertainty relation (\ref{2.15}) now
reads\,:
\begin{eqnarray}
\hat{\rho}\geq 0 \,\,\Rightarrow \,\,V(\hat{\rho}\,;\,\hat{A})+
\frac{i}{2}\,\omega(\hat{\rho}\,;\,\hat{A}) \geq 0,
\label{2.19}
\end{eqnarray}
and then by Lemma 2 we have the possible further consequence\,:
\begin{eqnarray}
{\rm det}\,\omega(\hat{\rho}\,;\,\hat{A}) \not= 0 \Rightarrow
V(\hat{\rho}\,;\,\hat{A}) > 0\,.
\label{2.20}
\end{eqnarray}

\vskip0.2cm

\noindent
{\bf Remark}\.:  
In case the operators $\hat{A}_{a}$ commute pairwise, in any state 
$\hat{\rho}$ there is a `classical' joint probability distribution
over the sets of simultaneous eigenvalues of all the $\hat{A}_{a}$. In
such a case, the term $\omega$ in Eqs.\,(\ref{2.18},\,\ref{2.19})
vanishes identically, and the uncertainty relation (\ref{2.19}) is a
`classical' statement\,\cite{fine-jmp82}. Therefore in the general 
case a good name for
$\omega_{ab}(\hat{\rho}\,; \,\hat{A})$ is that it is the `commutator
correction' term.

\vskip0.2cm

It is instructive to appreciate that while the original definitions of
$\Omega$ and $\tilde{\Omega}$, starting from the operator sets
$\hat{\cal A}$ and $\hat{\Omega}$, make it essentially trivial to see
that they must be nonnegative, the form (\ref{2.19}) of the general
uncertainty relation gives prominence to the variance matrix
$V(\hat{\rho}\,;\,\hat{A})$. In addition, as seen earlier, the matrix
$\Omega$ does not directly deal with fluctuations. It is after the use
of Lemma 1 that we obtain the matrix $\tilde{\Omega}$ involving the
fluctuations.

From Eqs.\,(\ref{2.8}, \ref{2.17}), the effect of a unitary symmetry
transformation on the real matrices $V(\hat{\rho}\,;\, \hat{A})$ and
$\omega(\hat{\rho}\,;\, \hat{A})$ is seen to be\,:
\begin{eqnarray}
\hat{\rho}^{\, \prime}=\overline{U}\,\hat{\rho}\,\overline{U}^{\,-1}\,: &&~
V(\hat{\rho}^{\, \prime}\,;\,\hat{A})=R\,V(\hat{\rho}\,;\, \hat{A})\, R^{T},
\nonumber \\
&&~\omega(\hat{\rho}^{\, \prime}\,;\,\hat{A})=R\,\omega(\hat{\rho}\,; \hat{A})\, R^{T},
\label{2.21}
\end{eqnarray}
so that the form (\ref{2.19}) of the uncertainty relation is manifestly
preserved. 

In later work, when there is no danger of confusion, we
sometimes omit the arguments $\hat{\rho}$ and $\hat{A}$ in $V$ and
$\omega$. 

\section{The multi mode Schr\"{o}dinger-Robertson Uncertainty Principle}
As a first example of the general framework we consider briefly the
Schr\"{o}dinger-Robertson UP for an $n$-mode system, which has been extensively
discussed elsewhere\,\cite{dutta94,gaussians}.

The basic operators, Cartesian coordinates and momenta, consist of
$n$ pairs of canonical $\hat{q}$ and $\hat{p}$ variables obeying the
Heisenberg canonical commutation relations. The operator properties
and relations are\,:
\begin{eqnarray}
a=1,\,2,\,\cdots,\,2n\,: \;\; \hat{\xi}_{a} &=& \left\{ \begin{array}{cc}
\hat{q}_{(a+1)/2}, \;& a ~ {\rm odd}\,, \\
\hat{p}_{a/2},\; & a ~ {\rm even}\,;
\end{array} \right. \nonumber \\
\hat{\xi}_{a}^{\dagger} &=& \hat{\xi}_{a} \,; \nonumber \\
\left[ \hat{\xi}_{a},\,\hat{\xi}_{b} \right] &=& i \hbar \beta_{ab}\,,\nonumber\\
\beta = \,{\rm block ~ diag}\,(\,i\sigma_2,\,i\sigma_2,\,\cdots,\,i\sigma_2\,) &=& 
{1\!\!1}_{n \times n} \otimes i\sigma_2\,.
\label{3.1}
\end{eqnarray}
These operators act irreducibly on the system Hilbert space ${\cal
  H}=L^{2}(\mathbb{R}^n)$. 

We take these $\hat{\xi}_{a}$ as the
$\hat{A}_{a}$ of Eq.\,(\ref{2.3}), so here $N=2n$\,:
\begin{eqnarray}
\hat{{\cal A}} \rightarrow 
\left( \begin{array}{c} 1 \\ \hat{\xi} \end{array} \right),\,\,\,
\hat{A}\rightarrow \hat{\xi}
    =\left( \begin{array}{c}
\hat{\xi}_{1} \\ \vdots \\\hat{\xi}_{2n} \end{array}\right)
    =\left( \begin{array}{c}
\hat{q}_{1} \\ \hat{p}_{1}\\ \vdots \\ \hat{q}_{n} \\ \hat{p}_{n} \end{array}\right).
\label{3.2}
\end{eqnarray}
Then for any state $\hat{\rho}$, the variance matrix $V$ has elements
\begin{eqnarray}
V_{ab}&=& \frac{1}{2}\,{\rm Tr} \left(\hat{\rho}\,\{\hat{\xi}_{a}-{\rm
  Tr}(\hat{\rho}\,\hat{\xi}_{a}),\, \hat{\xi}_{b} - {\rm
  Tr}(\hat{\rho}\,\hat{\xi}_{b}) \}\right) \nonumber \\
&=&\frac{1}{2}\langle\{ \hat{\xi}_{a},\, \hat{\xi}_{b}\} \rangle -
\langle \hat{\xi}_{a} \rangle\,\langle \hat{\xi}_{b} \rangle,
\label{3.3}
\end{eqnarray}
while the antisymmetric matrix $\omega$ is just the {\em state-independent}
numerical `symplectic metric matrix' $\beta$\,: 
\begin{eqnarray}
\omega_{ab}= -i \, \langle\left[\hat{\xi}_{a},\,\hat{\xi}_{b}
  \right] \rangle = \hbar \beta_{ab}.
\label{3.4}
\end{eqnarray}
The uncertainty relation (\ref{2.19}) then becomes the $n$-mode
Schr\"{o}dinger-Robertson UP\,:
\begin{eqnarray}
\hat{\rho}\geq 0 \Rightarrow V + i \frac{\hbar}{2} \beta \geq
0\,\,(\Rightarrow V > 0),
\label{3.5}
\end{eqnarray}
the second step following from Eq.\,(\ref{2.20}) as $\beta$ is nonsingular.

For $n=1$, a single mode, the matrices $V$ and $\beta$ are
two-dimensional\,:
\begin{eqnarray}
&& V = \left( \begin{array}{cc}
(\Delta q)^2 & \Delta (q,p) \\
\Delta (q,p) & (\Delta p)^2
\end{array} \right), \nonumber \\
&& (\Delta q)^2 = \langle (\hat{q} -\langle \hat{q} \rangle)^2
  \rangle,\,\,\,\,
(\Delta p)^2 = \langle (\hat{p} -\langle \hat{p} \rangle)^2\rangle, \nonumber
  \\
&& \Delta (q,p) = \frac{1}{2}\langle \{\hat{q}-\langle \hat{q} \rangle,\,
  \hat{p} - \langle \hat{p} \rangle \} \rangle\,; \nonumber \\ 
&& \beta = i \sigma_{2} = \left(\begin{array}{cc} 0 & 1 \\ -1 & 0 
\end{array} \right).
\label{3.6}
\end{eqnarray}
Then (\ref{3.5}) simplifies to
\begin{eqnarray}
\left( \begin{array}{cc}
(\Delta q)^2 & \Delta (q,p)+\frac{i}{2} \hbar \\
\Delta (q,p)-\frac{i}{2}\hbar & (\Delta p)^2
\end{array} \right) &\geq& 0,\nonumber\\
i.e., \; {\rm det}\left( V + \frac{i}{2}\hbar \beta \right)\equiv
(\Delta q)^2 (\Delta p)^2 -(\Delta(q,p))^2 -\frac{\hbar^2}{4} &\geq& 0,\nonumber\\
i.e., \; {\rm det}\,V  &\geq& \frac{\hbar^2}{4}\,,
\label{3.7}
\end{eqnarray}
the original Schr\"{o}dinger-Robertson UP. 

Returning to $n$ modes, the $Sp(2n,\,R)$
covariance of the Schr\"{o}dinger-Robertson UP (\ref{3.5}) takes the following 
form\,:
If $S \in Sp(2n,\, R)$, i.e., any real $2n \times 2n$ matrix obeying $S
\beta S^T = \beta$, then the new operators
\begin{eqnarray}
\hat{\xi}_{a}^{'}=S_{ab} \, \hat{\xi}_{b}
\label{3.8}
\end{eqnarray}
preserve the commutation relations in Eq.\,(\ref{3.1}) and hence are
unitarily related to the $\hat{\xi}_{a}$. These unitary
transformations constitute the double valued metaplectic unitary
representation of $Sp(2n,\,R)$\,\cite{pramana95}\,:
\begin{eqnarray}
S \in Sp(2n,\,R) &\rightarrow& \overline{U}(S)={\rm unitary\,\,operator\,\,
  on\,\,{\cal H}}, \nonumber \\
\overline{U}(S^{\, \prime})\overline{U}(S) &=& \pm \overline{U}(S^{\, \prime} S)\,; \nonumber \\
\overline{U}(S)^{-1}\,\hat{\xi}_{a}\, \overline{U}(S)&=& S_{ab}\,\hat{\xi}_{b}.
\label{3.9}
\end{eqnarray}
Then, as an instance of Eqs.\,({\ref{2.21}}) we have the results\,:
\begin{eqnarray}
&&\hat{\rho}\rightarrow \hat{\rho}^{\, \prime} =\overline{U}(S)\,\hat{\rho}\,
\overline{U}(S)^{-1} \Rightarrow V \rightarrow V^{\, \prime} = S\,V \, S^{T},
\nonumber \\
&&V + \frac{i}{2}\,\hbar\, \beta \geq 0 \,\, \Leftrightarrow \,\, V^{\, \prime} +
\frac{i}{2}\, \hbar\, \beta \geq 0.
\label{3.10}
\end{eqnarray}

\vskip0.2cm

\noindent
{\bf Remark}\,: 
The $n$-mode Schr\"{o}dinger-Robertson UP (\ref{3.5}), with its
explicit $Sp(2n,\,R)$ covariance (\ref{3.10}), constitutes the answer
to an important question raised by Littlejohn\,\cite{littlejohn86}: under what conditions
is a real normalized Gaussian function on a $2n$-dimensional phase space
the Wigner distribution for some quantum state? The answer is stated
in terms of the variance matrix which of course determines the
Gaussian up to phase space displacements [And these phase space displacements 
have no role to play on the `Wigner quality' of a phase space distribution]. 
 This result has been used
extensively in both classical and quantum optics\,\cite{gaussians}, and more recently 
in quantum information theory of continuous variable canonical systems\,\cite{cv}. 

\vskip0.2cm

As a last comment we mention that as according to Eq.\,(\ref{3.5}) the
variance matrix $V$ is always positive definite, by Williamson's
celebrated theorem an $S \in Sp(2n,\,R)$ can be found such that $V^{\, \prime}$
in Eq.\,(\ref{3.10}) becomes diagonal\,\cite{Williamson,RSSCVS}. In general, 
though, the diagonal elements of $V^{\, \prime}$ will not be the eigenvalues of $V$.

\section{Higher order moments for single mode system}
We now revert to the $n=1$ case of one canonical pair of hermitian
operators $\hat{q}$ and $\hat{p}$, but consider expectation values of
expressions in these operators of order greater than two. The relevant
Hilbert space is of course ${\cal H}= L^{2}(\mathbb{R})$. As a useful
computational tool we work with the Wigner distribution description of
quantum states, and the associated Weyl rule of association of
(hermitian) operators with (real) classical phase space functions.

Given a quantum mechanical state $\hat{\rho}$, the corresponding
Wigner distribution is a function on the classical two-dimensional
phase space\,:
\begin{eqnarray}
W(q, p)=\frac{1}{2 \pi \hbar} \int_{-\infty}^{\infty} dq^{\, \prime} \left\langle
q -\frac{1}{2} q^{\, \prime} \right|\hat{\rho} \left| q + \frac{1}{2} q^{\, \prime} \right\rangle
e^{ipq^{\, \prime}/\hbar}.
\label{4.1}
\end{eqnarray}
Thus it is a partial Fourier transform of the position space matrix
elements of $\hat{\rho}$. This function is real and normalised to unity,
but need not be pointwise nonnegative\,:
\begin{eqnarray}
\hat{\rho}^{\dagger}=\hat{\rho} & \Rightarrow &  W(q,p)^{*}=W(q,p)\,;
  \nonumber \\
{\rm Tr}\,\hat{\rho}=1 &\Rightarrow&
\int_{-\infty}^{\infty}\int_{-\infty}^{\infty}dqdp \,W(q,p)=1.
\label{4.2}
\end{eqnarray}
The operator $\hat{\rho}$ and the function $W(q,p)$ determine each
other uniquely. The {\em key property} is that the quantum expectation
values of operator exponentials are equal to the classical phase space
averages of classical exponentials with respect to $W(q,p)$
\,\cite{Cahill}\,:
\begin{eqnarray}
{\rm Tr}(\hat{\rho}\,e^{i(\theta\, \hat{q}- \tau\, \hat{p})}) =
\int_{-\infty}^{\infty}\int_{-\infty}^{\infty} dq dp \,W(q,p) 
e^{i(\theta\, {q}- \tau\, {p})},\,\,\,-\infty <\, \theta,\, \tau\,<
\infty. 
\label{4.3}
\end{eqnarray}
By expanding the exponentials and comparing powers of $\theta$ and
$\tau$ we get\,:
\begin{eqnarray}
{\rm Tr}(\hat{\rho}\, \widehat{(q^n\,p^{n'})}) &=&
\int_{-\infty}^{\infty} \int_{-\infty}^{\infty} dq dp \,W(q,p)
q^{n}p^{n'}, \nonumber \\
\widehat{(q^n\,p^{n'})}&=&{\rm coefficient\,\,of\,\,}
\frac{(i\theta)^n}{n!}\,\frac{(-i\tau)^{n'}}{{n'}!}\,\,{\rm in}\,\,\
e^{i(\theta\, \hat{q}- \tau\, \hat{p})} \nonumber \\
&=&\frac{n!\,{n'}!}{(n+n')!} \times \,\,{\rm coefficient \,\,of \,\,}
\theta^n(-\tau)^{n'} \,\,{\rm in}\,\,(\theta\,\hat{q}-
\tau\,\hat{p})^{n + n'}, \nonumber \\
&& n,\,n'=0,\,1,\,2,\, \cdots. 
\label{4.4}
\end{eqnarray}
Thus $\widehat{(q^n\,p^{n'})}$ is an hermitian operator polynomial in
$\hat{q}$ and $\hat{p}$ associated to the classical real monomial
$q^np^{n'}$. This is the Weyl rule of association indicated by 
\begin{eqnarray}
\widehat{(q^n\,p^{n'})}={(q^n\,p^{n'})}_{W},
\label{4.5}
\end{eqnarray}
so Eq.\,(\ref{4.4}) appears as 
\begin{eqnarray}
{\rm Tr}(\hat{\rho}\,{(q^n\,p^{n'})}_{W} )=
\int \int dq dp\, W(q,p) \,q^n p^{n'}.
\label{4.6}
\end{eqnarray}
We regard the polynomials  ${(q^n\,p^{n'})}_{W}$ as the basic `quantum
monomials'. By linearity the association (\ref{4.5}) can be extended to
general functions on the classical phase space, leading to the
scheme\,:
\begin{eqnarray}
f(q,p)={\rm real\,\,classical\,\,function\,\,} &\rightarrow & \hat{F}=
(f(q,p))_{W}={\rm hermitian\,\, operator\,\, on \,\,{\cal H}},
\nonumber \\
{\rm Tr}(\hat{\rho}\,\hat{F})&=& \int \int dqdp\, W(q,p)\,f(q,p).
\label{4.7}
\end{eqnarray}

\vskip 0.2cm
\noindent
{\bf Remarks}: 
Two useful comments may be made at this point. For any pair of states
$\hat{\rho},\;\hat{\rho}^{\prime}$ we have 
\begin{eqnarray}
{\rm Tr}(\hat{\rho}\,\hat{\rho}^{\prime})&=& \int \int dq dp\, 
W(q,p)\,W^{\prime}(q,p)\geq 0.
\label{4.7a}
\end{eqnarray}
Based on this, one can see the following\,: a given real normalised
phase space function $W(q,p)$ is a Wigner distribution (corresponding
to some physical state $\hat{\rho}$) if and only if the overlap integral on
the right hand side of Eq.\,(\ref{4.7a}) is nonnegative for all Wigner
distributions $W^{\, \prime}(q,p)$. Secondly, we refer to the remarks made
following Eq.\,(\ref{2.20}) concerning the commutative case
$\left[\hat{A}_{a},\,\hat{A}_{b} \right]=0$. This happens for
instance when $\hat{A}_{a}=f_{a}(\hat{q})$ for all $a$. In that case,
only the integral of $W(q,p)$ over $p$ is relevant, and this is known
to be the coordinate space probability density in the state
$\hat{\rho}$\,\cite{hilleryPR}. In the multi mode case this generalizes to the following
statement\,: the result of integrating
$W(q_1,p_1,\,q_2,p_2,\,\cdots\,,\,q_n,p_n)$ over any
($n$-dimensional) linear Lagrangian subspace in phase space is always a genuine 
probability distribution (the marginal) over the `remaining' $n$ phase space
variables. {\em This marginal is basically the squared modulus, or probability density 
 in the Born sense, of a wavefunction on the corresponding `configuration space', 
 generalised to the case of a mixed state}\,\cite{hilleryPR}.  
 
The covariance group  of the
canonical commutation relation obeyed by $\hat{q}$ and $\hat{p}$ 
 is (apart from phase space translations) the group $Sp(2,\,R)$\,:
\begin{eqnarray}
Sp(2,\,R)&=& \left\{ S=\left(\begin{array}{cc} a & b \\ c &
  d \end{array} \right) ={\rm real \,\,} 2\times 2\,\,{\rm matrix}
\,\,|\,\,S\,\sigma_2\,S^{T} =\sigma_2,\,\;{\rm i.e.,}\; {\rm det}\,S =1 
\right\}\,.~~
\label{4.8}      
\end{eqnarray}
The actions on $\hat{q}$ and $\hat{p}$ by matrices and by the unitary
metaplectic representation of $Sp(2,\,R)$ are connected in this
manner\,:
\begin{eqnarray}
S \in Sp(2,\,R) &\rightarrow &\overline{U}(S)={\rm
  unitary\,\,operator\,\,on\,\,{\cal H}}\,; \nonumber \\
\xi = \left(\begin{array}{c} q \\ p \end{array} \right)
& \rightarrow & \hat{\xi} =(\xi)_{W}= \left(\begin{array}{c} 
\hat{q} \\ \hat{p} \end{array} \right)\,: \nonumber \\
\overline{U}(S)^{-1}\,\hat{\xi}\,\overline{U}(S)&=& S\,\hat{\xi}.
\label{4.9}
\end{eqnarray}
The effect on $W(q,p)\equiv W(\xi)$ is then given as\,\,\cite{dutta94,gaussians}\,:
\begin{eqnarray}
\hat{\rho}^{\, \prime}=\overline{U}(S)\,\hat{\rho}\,\overline{U}(S)^{-1} \leftrightarrow
W^{\, \prime}(\xi)=W(S^{-1}\,\xi).
\label{4.10}
\end{eqnarray}

We now introduce a more suggestive notation for the classical
monomials $q^n p^{n'}$ and their operator counterparts $(q^n
p^{n'})_{W}$. This is taken from the quantum theory of angular
momentum (QTAM) and uses the fact that finite-dimensional nonunitary
irreducible real representations of $Sp(2,\,R)$ are related to the
unitary irreducible representations of $SU(2)$ by analytic
continuation. (Indeed the two sets of generators are related by the
unitary Weyl trick). We use `quantum numbers' $j=0,\,
\frac{1}{2},\,1,\, \cdots$, $m = j,\,j-1,\, \cdots,\, -j$ as in QTAM
and define the hermitian monomial basis for operators on ${\cal H}$ in
this way\,:
\begin{eqnarray}
\hat{T}_{jm}=(q^{j+m}p^{j-m})_{W} &=& {\rm coefficient \,\,of\,\,}
\frac{(2j)!}{(j+m)!(j-m)!} \,\theta^{j+m} (-\tau)^{j-m}\,\,{\rm in}
\,\, (\theta\,\hat{q}-\tau\,\hat{p})^{2j},\nonumber \\
&&j=0,\,\frac{1}{2},\,1,\,\cdots,\,;\,\,\,m=j,\,j-1,\,\cdots,\, -j.
\label{4.11}
\end{eqnarray}
For the first few values of $j$ we have
\begin{eqnarray}
&&(\hat{T}_{\frac{1}{2} m})= \left(\begin{array}{c} \hat{q}\\ \hat{p}
\end{array} \right);\,\,\,
(\hat{T}_{1 m})= \left(\begin{array}{c} \hat{q}^2\\ 
\frac{1}{2}\{\hat{q},\, \hat{p}\} \\ \hat{p}^2
\end{array} \right);\,\,\,
(\hat{T}_{\frac{3}{2} m})= \left(\begin{array}{c} \hat{q}^3\\ 
\frac{1}{3}(\hat{q}^2\hat{p} + \hat{q}\hat{p}\hat{q}+ \hat{p}
\hat{q}^2)\\
\frac{1}{3}(\hat{q}\hat{p}^2 + \hat{p}\hat{q}\hat{p}+ \hat{p}^2
\hat{q})\\
\hat{p}^3
\end{array} \right); \nonumber \\
&&(\hat{T}_{2 m})= \left(\begin{array}{c} \hat{q}^4\\ 
\frac{1}{4}(\hat{q}^3\hat{p} + \hat{q}^2\hat{p}\hat{q}+ \hat{q}\hat{p}\hat{q}^2
+ \hat{p} \hat{q}^3)\\
\frac{1}{6}(\hat{q}^2 \hat{p}^2 +\hat{q}\hat{p}\hat{q}\hat{p}
+\hat{q}\hat{p}^2 \hat{q} + \hat{p}\hat{q}^2 \hat{p} +
\hat{p}\hat{q}\hat{p}\hat{q} + \hat{p}^2 \hat{q}^2) \\
\frac{1}{4}(\hat{q}\hat{p}^3 + \hat{p}\hat{q}\hat{p}^2+ \hat{p}^2\hat{q}\hat{p}
+ \hat{p}^3\hat{q})\\
\hat{p}^4
\end{array} \right).
\label{4.12}
\end{eqnarray} 
Then we have the consequences\,:
\begin{eqnarray}
&& {\rm Tr}(\hat{\rho}\,\hat{T}_{jm}) = \int \int dqdp\, W(q,p)
\,q^{j+m}\,p^{j-m} \equiv \overline{q^{j+m}\,p^{j-m}}\,; \nonumber \\
&& S\in Sp(2,\,R):\;\, \overline{U}(S)^{-1}\, \hat{T}_{jm}\, \overline{U}(S) =
\sum_{m'= -j}^{j} K^{(j)}_{m m'} (S)\,\hat{T}_{jm'}.
\label{4.13}
\end{eqnarray}
The quantum expectation values of the $\hat{T}_{jm}$ are phase space
moments of $W(q,p)$, denoted for convenience with an overhead bar.
The matrices $K^{(j)}(S)$ constitute the $(2j+1)$-dimensional real
nonunitary irreducible representation of $Sp(2,\,R)$ obtained from the
familiar `spin $j$' unitary irreducible representation of $SU(2)$ by
analytic continuation. For $j = \frac{1}{2}$, we have
$K^{(1/2)}(S)=S$. The representation $K^{(1)}(S)$ corresponding to $j=1$ will 
be seen to engage our sole attention in Section~V.

The noncommutative (but associative) product law for
the hermitian monomial operators $\hat{T}_{jm}$ has an interesting
form, being essentially determined by the $SU(2)$ Clebsch-Gordan
coefficients.  
This is not surprising, in view of the connection between $SU(2)$ and
$Sp(2,\,R)$ representations (in finite dimensions) mentioned
above. In fact for these representations and in chosen bases, $SU(2)$ and
$Sp(2,\, R)$ share the same Clebsch-Gordan coefficients\,\cite{moments2012}. The product
formula has a particularly simple structure if we (momentarily) use
suitable numerical multiples of $\hat{T}_{jm}$\,:
\begin{eqnarray}
\hat{\tau}_{jm}= \hat{T}_{jm}/\sqrt{(j+m)!(j-m)!}.
\label{4.14}
\end{eqnarray} 
Then we find\,\cite{moments2012}
\begin{eqnarray}
\hat{\tau}_{jm}\,\hat{\tau}_{j' m'} &=& \sum _{j'' = |j-j'|}^{j+j'}
\left(\frac{i \hbar}{2} \right)^{j+j'-j''}
\sqrt{\frac{(j+j' + j'' +1)!}{(2j'' +1)(j+j'-j'')! (j'+j'' -j)! (j'' +j
  -j')!}} \nonumber \\
&& \times C^{j\,j'\,j''}_{m\,m'\, m+m'}\, \hat{\tau}_{j'',\,m +m'}.
\label{4.15}
\end{eqnarray}
The $C^{j\,j'\,j''}_{m\,m'\,m''}$ are the $SU(2)$ Clebsch-Gordan
coefficients familiar from QTAM\,\cite{edmonds}. We will use this 
product rule in the sequel.

Now we apply the general framework of Section II to the present
situation. We will use a notation similar to that in the main theorems
of the classical theory of moments. We take $\hat{A}$ and $\hat{{\cal A}}$
to formally be infinite component column vectors with hermitian
entries\,:
\begin{eqnarray}
\hat{A} &=& \left(\begin{array}{c}  \vdots \\ 
\hat{T}_{jm} \\ \vdots \end{array} \right) =
(\hat{T}_{\frac{1}{2}\, \frac{1}{2}}, \,\hat{T}_{\frac{1}{2}\,
  \frac{-1}{2}},\, \hat{T}_{1\,1},\, \hat{T}_{1\,0},\,\hat{T}_{1\,
  -1},\,\cdots, \,\hat{T}_{jj},\,\cdots,\,
\hat{T}_{j,-j},\,\cdots)^{T}, \nonumber \\
\hat{\cal A}&=& \left(\begin{array}{c}1 \\ \hat{A} \end{array} \right).
\label{4.16}
\end{eqnarray}
Thus the subscript $a$ of Eq.\,(\ref{2.4}) is now the pair $jm$ taking
values in the sequence given above. To simplify notation, as $\hat{A}$
is kept fixed, we will not indicate it as an argument in various
quantities. The general entries in the infinite-dimensional matrices 
$\hat{\Omega}$, $\Omega$, $\tilde{\Omega}$ in Eqs.\,(\ref{2.4},\,\,\ref{2.5})
are then\,:
\begin{eqnarray}
\hat{\Omega}_{jm, j'm'} &=& \hat{T}_{jm}\,\hat{T}_{j'm'}\,; \nonumber \\
\Omega_{jm,j'm'}(\hat{\rho}) &=&{\rm
  Tr}(\hat{\rho}\,\hat{T}_{jm}\,\hat{T}_{j'm'})=
\langle \hat{T}_{jm}\,\hat{T}_{j'm'} \rangle\,; \nonumber \\
\tilde{\Omega}_{jm,j'm'}(\hat{\rho}) &=&
\langle (\hat{T}_{jm}- \langle\hat{T}_{jm}
\rangle)\,(\hat{T}_{j'm'}-\langle \hat{T}_{j'm'} \rangle) \rangle.
\label{4.17}
\end{eqnarray}
(In $\hat{\Omega}$ and $\Omega$, for $j=m=0$, we have $\hat{T}_{00}
=1$). By using the product rule (\ref{4.15}) the (generally nonhermitian)
operator $\hat{T}_{jm}\,\hat{T}_{j'm'}$ can be written as a complex
linear combination of $\hat{T}_{j'',m+m'}$ with $j'' = j+j',\,
j+j'-1,\,\cdots,\, |j-j'|$. The variance matrix $V(\hat{\rho})$ in
Eq.\,(\ref{2.18}) has the elements
\begin{eqnarray}
V_{jm,j'm'}(\hat{\rho}) =\frac{1}{2} \langle \{\hat{T}_{jm},
\,\hat{T}_{j'm'} \} \rangle - \langle \hat{T}_{jm} \rangle\,\langle
\hat{T}_{j'm'} \rangle.  
\label{4.18}
\end{eqnarray}
From the known symmetry relation \,\cite{edmonds}
\begin{eqnarray}
C^{j'\,j\,j''}_{m'\,m\, m+m'} = (-1)^{j+j'-j''}\,C^{j\,j'\,j''}_{m\,m'\,m+m'}
\label{4.19}
\end{eqnarray}
we see that in the anticommutator term in Eq.\,(\ref{4.18}) only
$\hat{T}_{m+m'}^{j''}$ for $j''= j+j',\,j+j'-2,\, j+j'-4, \, \cdots$
will appear with real coefficients. On the other hand, for the
antisymmetric part $\omega_{ab}$ of Eq.\,(\ref{2.18}) we have
\begin{eqnarray}
\omega_{jm,j'm'}(\hat{\rho}) = -i \left\langle \left[\hat{T}_{jm}
  ,\,\hat{T}_{j'm'}\right] \right\rangle ,
\label{4.20}
\end{eqnarray}
so now by Eq.\,(\ref{4.19}) the commutator here is a linear combination of
terms $\hat{T}_{m+m'}^{j''}$ for $j''= j+j'-1,\,j+j'-3,\, \cdots$ with
pure imaginary coefficients. There is, therefore, a clean separation
of the product $\hat{T}_{jm}\, \hat{T}_{j'm'}$ into a hermitian part
in $V$ and an antihermitian part in $\omega$. With these facts in
mind, the uncertainty relation (\ref{2.19}) is in hand\,:
\begin{eqnarray}
V_{jm,j'm'}(\hat{\rho})&=&\sum_{j+j'-j''\,\,{\rm even}}
\cdots\,\,\langle \hat{T}_{j'',m+m'} \rangle - \langle \hat{T}_{jm}
\rangle\, \langle \hat{T}_{j'm'} \rangle, \nonumber \\
\omega_{jm,j'm'}(\hat{\rho})&=&  \sum_{j+j'-j''\,\,{\rm odd}}
\cdots\,\,\langle \hat{T}_{j'',m+m'} \rangle\,; \nonumber \\
(\tilde{\Omega}_{jm,j'm'}(\hat{\rho})) &=&(V_{jm,j'm'}(\hat{\rho}))
+\frac{i}{2}\,(\omega_{jm,j'm'}(\hat{\rho})) \geq 0.
\label{4.21}
\end{eqnarray}
Each matrix element of $V(\hat{\rho})$ (apart from the subtracted
term) and of $\omega(\rho)$ appears as some real linear combination
of expectation values of hermitian monomial operators, i.e., of
moments of $W(q,p)$\,; however, in this way of writing, the
essentially trivial nature of the statement
$\tilde{\Omega}(\hat{\rho}) \geq 0$ is not manifest.

The covariance group in this problem is of course $Sp(2,\,R)$. From
Eq.\,(\ref{4.13}) we see that under conjugation by the metaplectic group
unitary operator $\overline{U}(S)$, the column vector $\hat{A}$ of
Eq.\,(\ref{4.16}) transforms as a direct sum of the sequence of finite-dimensional real irreducible nonunitary representation matrices 
$K^{(1/2)}(S)=S$, $K^{(1)}(S)$, $K^{(3/2)}(S)\,\cdots$\,; so
Eq.\,(\ref{2.8}) in the present context is\,\cite{moments2012}\,:
\begin{eqnarray}
&&S\in Sp(2,\,R)\,:\,\,\,\, \overline{U}(S)^{-1}\, \hat{A}\, \overline{U}(S)
= K(S)\,\hat{A}, \nonumber \\
&&K(S)= K^{(1/2)}(S)\oplus K^{(1)}(S)\oplus K^{(3/2)}(S)\,\oplus \,\cdots
\label{4.22}
\end{eqnarray}
From Eq.\,(\ref{2.21}), when $\hat{\rho}\rightarrow \hat{\rho}'=
\overline{U}(S) \,\hat{\rho}\, \overline{U}(S)^{-1}$ both $V(\hat{\rho})$ and 
$\omega(\hat{\rho})$ experience congruence transformations by $K(S)$,
and the formal uncertainty relation (\ref{4.21}) is preserved.

Up to this point the use of infinite component $\hat{A}$ and infinite-dimensional 
$\Omega$, $\tilde{\Omega}$, $V$ and $\omega$ has been
formal. We may now interpret the uncertainty relation (\ref{4.21}) in
practical terms to mean that for each finite $N=1,\,2,\,\cdots\,$, the
principal submatrix of $\tilde{\Omega}(\hat{\rho})$ formed by its
first $N$ rows and columns should be nonnegative. However, in order to
maintain $Sp(2,\,R)$ covariance, a slight modification of this
procedure is desirable. If for each
$J=\frac{1}{2},\,1,\,\frac{3}{2},\,\cdots$ we include all values of
$j\,m$ for $j\leq J$, the number of rows (and columns) involved is
$N_{J}=J(2J+3)$, the sequence of integers $2,\,5,\,9,\,14,\,\cdots$.
Let us then define hierarchies of $N_{J}$-dimensional matrices as\,:
\begin{eqnarray}
J=\frac{1}{2},\,1,\,\frac{3}{2},\,\cdots\,:~~&& \nonumber\\
\tilde{\Omega}^{(J)}(\hat{\rho})&=&
(\tilde{\Omega}_{jm,j'm'}(\hat{\rho})), \nonumber \\
V^{(J)}(\hat{\rho})   &=&
(V_{jm,j'm'}(\hat{\rho})), \nonumber \\
{\omega}^{(J)}(\hat{\rho}) &=&
({\omega}_{jm,j'm'}(\hat{\rho})), 
\,\,\,j,\,j'=\frac{1}{2},\,1,\,\cdots,\,J\,;
\nonumber \\
\tilde{\Omega}^{(J)}(\hat{\rho}) &=&
V^{(J)}(\hat{\rho})+\frac{i}{2}
{\omega}^{(J)}(\hat{\rho}).
\label{4.23}
\end{eqnarray}
However, in each of these matrices {\em there is no $J$
  dependence in their matrix elements}. Each also naturally breaks up 
into blocks of dimension $(2j+1)(2j^{\, \prime}+1)$ for each pair $(j,j^{\, \prime})$
present, and these can be denoted by
$\tilde{\Omega}^{(j,j')}(\hat{\rho})$, $V^{(j,j')}(\hat{\rho})$, 
${\omega}^{(j,j')}(\hat{\rho})$. Symbolically,
\begin{eqnarray}
\tilde{\Omega}^{(J)}(\hat{\rho})= \left( 
\begin{array}{ccc}
& \vdots &  \\
\cdots & \tilde{\Omega}^{(j,j')}(\hat{\rho}) & \cdots \\
& \vdots & 
\end{array}
\right)
\label{4.24}
\end{eqnarray}
and similary for $V^{(J)}(\hat{\rho})$ and
$\omega^{(J)}(\hat{\rho})$. As examples we have\,:
\begin{eqnarray}
\tilde{\Omega}^{(1/2)}(\hat{\rho})&=&(\tilde{\Omega}^{\left(
  \frac{1}{2}, \frac{1}{2} \right)}(\hat{\rho}))\,; \nonumber \\
\tilde{\Omega}^{(1)}(\hat{\rho})&=&
\left(\begin{array}{cc}
\tilde{\Omega}^{\left(
  \frac{1}{2}, \frac{1}{2} \right)}(\hat{\rho}) &
\tilde{\Omega}^{\left(
  \frac{1}{2}, 1 \right)}(\hat{\rho})\\
\tilde{\Omega}^{\left(
  1 , \frac{1}{2} \right)}(\hat{\rho}) &
\tilde{\Omega}^{\left(
 1,1 \right)}(\hat{\rho})
\end{array} \right),
\label{4.25}
\end{eqnarray}
and correspondingly for $V^{(J)}$, $\omega^{(J)}$. Moreover, in going
from $J$ to $J +\frac{1}{2}$, we have an augmentation of each matrix
with $2(J+1)$ new rows and columns,
\begin{eqnarray}
\tilde{\Omega}^{(J + 1/2)}(\hat{\rho})=
\left( \begin{array}{ccc}
\tilde{\Omega}^{(J)}(\hat{\rho})  & 
\begin{array}{c}
\vdots \\
\vdots \\
\vdots 
\end{array}
& \begin{array}{c}
\tilde{\Omega}^{\left(
  \frac{1}{2}, J + \frac{1}{2} \right)}(\hat{\rho}) \\
\vdots  \\
\tilde{\Omega}^{\left(
  J, J+ \frac{1}{2} \right)}(\hat{\rho}) 
\end{array} \\
\begin{array}{ccccc}
\cdots\,\,\,\, &\,\,\,\, \cdots\,\,\,\, &\,\,\,\, \cdots\,\,\,\, &
\,\,\,\,\cdots\,\,\,\,& \,\,\,\,\cdots 
\end{array} && \cdots\,\,\,\,\cdots \\
\begin{array}{ccc}
\tilde{\Omega}^{\left(
  J+ \frac{1}{2}, \frac{1}{2} \right)}(\hat{\rho}) & \cdots & \tilde{\Omega}^{\left(
  J + \frac{1}{2}, J \right)}(\hat{\rho}) 
\end{array} &\vdots & \tilde{\Omega}^{\left(
 J+ \frac{1}{2}, J+ \frac{1}{2} \right)}(\hat{\rho})  
 \end{array}\right).
\label{4.26}
\end{eqnarray}
The formal uncertainty relation (\ref{4.21}) now translates into a
hierarchy of finite-dimensional matrix conditions
\begin{eqnarray}
\tilde{\Omega}^{(J )}(\hat{\rho}) =
V^{(J )}(\hat{\rho})+ \frac{i}{2}
{\omega}^{(J)}(\hat{\rho}) \geq 0,\,\,J=\frac{1}{2},\,1,\,
\frac{3}{2},\, \cdots.
\label{4.27}
\end{eqnarray}
(Of course, for a given state $\hat{\rho}$, moments may exist and
be finite only up to some value $J_{\rm max}$ of $J$, so the hierarchy 
(\ref{4.27}) also terminates at this point).
 The lowest condition in this hierarchy, $J=\frac{1}{2}$,  takes us back to
Eqs.\,(\ref{3.6},\,\,\ref{3.7})\,:
\begin{eqnarray}
\tilde{\Omega}^{\left(\frac{1}{2} \right)}(\hat{\rho}) &=&
\tilde{\Omega}^{\left(
\frac{1}{2}, \frac{1}{2} \right)}(\hat{\rho})=
V^{\left(\frac{1}{2} \right)}(\hat{\rho})+ \frac{i}{2}
 {\omega}^{\left( \frac{1}{2} \right)}(\hat{\rho})\,;
\nonumber \\
 V^{\left(\frac{1}{2} \right)}(\hat{\rho}) &=&
V^{\left(\frac{1}{2}, \frac{1}{2} \right)}(\hat{\rho}) =
\left( \left\langle \frac{1}{2} \{\hat{T}_{\frac{1}{2}, m},
\,\hat{T}_{\frac{1}{2}, m'} \}\right\rangle\right)-
\left( \begin{array}{c}
\langle \hat{q} \rangle \\
\langle \hat{p} \rangle
\end{array}
\right)\,
\begin{array}{c}
\left( \begin{array}{cc}
\langle\hat{q}\rangle &\langle\hat{p} \rangle
\end{array} \right)
\\
\\
\end{array} \nonumber \\
&=& \left( \begin{array}{cc}
(\Delta q)^2 & \Delta(q,p) \\
\Delta (q,p) & (\Delta p)^2
\end{array} \right)\,; \nonumber \\
{\omega}^{\left(\frac{1}{2} \right)}(\hat{\rho}) &=&
{\omega}^{\left(
\frac{1}{2}, \frac{1}{2} \right)}(\hat{\rho})=
-i \left( \begin{array}{cc}
0 & \left[\hat{q},\,\hat{p} \right] \\
\left[\hat{p},\,\hat{q} \right] & 0
\end{array} \right)= i\,\hbar\, \sigma_2\,; \nonumber \\
\tilde{\Omega}^{\left(\frac{1}{2} \right)}(\hat{\rho}) &\geq&
0\,\,\,\,\,\Leftrightarrow \,\,\,\,\,
\left( \begin{array}{cc}
(\Delta q)^2 & \Delta(q,p)+ i \,\frac{\hbar}{2} \\
\Delta (q,p)-i\, \frac{\hbar}{2} & (\Delta p)^2
\end{array} \right)\,\ge 0\,\,\Leftrightarrow \nonumber \\
&& (\Delta q)^2\,(\Delta p)^2 - (\Delta (q,p))^2 \geq \frac{\hbar^2}{4},
\label{4.28}
\end{eqnarray}
the original Schr\"{o}dinger-Robertson  UP.

It is natural to ask for the new conditions that appear at each step in
the hierarchy (\ref{4.27}), in passing from $J$ to $J +\frac{1}{2}$. In
the generic case, when we have a strict inequality we can find the
answer using Lemma 1 of Section II. Comparing 
$\tilde{\Omega}^{\left(J+ \frac{1}{2} \right)}(\hat{\rho})$
and $\tilde{\Omega}^{\left(J\right)}(\hat{\rho})$, in the notation of
  Eq.\,(\ref{2.10}) and using Eq.\,(\ref{4.26}) we have\,:
\begin{eqnarray}
&&\tilde{\Omega}^{\left(J+ \frac{1}{2} \right)}(\hat{\rho})=
\left( \begin{array}{cc}
A & C^{\dagger} \\
C & B
\end{array} \right)\,;\nonumber \\
&& A = \tilde{\Omega}^{\left(J \right)}(\hat{\rho})\,,\,\,\,
\,\,B = \tilde{\Omega}^{\left(J+ \frac{1}{2}, J +\frac{1}{2}
  \right)}(\hat{\rho}), \nonumber \\
&& C= \left(  \begin{array}{ccc}
\tilde{\Omega}^{\left(J+ \frac{1}{2}, \frac{1}{2}
  \right)}(\hat{\rho})& \cdots &
\tilde{\Omega}^{\left(J+ \frac{1}{2}, J 
  \right)}(\hat{\rho})
\end{array} \right).
\label{4.29}
\end{eqnarray}
The `dimensions' are $N_{J} \times N_{J}$, $2(J+1)\times 2(J+1)$,
$2(J+1)\times N_{J}$ respectively. Then 
\begin{eqnarray}
\tilde{\Omega}^{\left(J +\frac{1}{2} \right)}(\hat{\rho}) > 0 \,\,
\Leftrightarrow \,\, \tilde{\Omega}^{\left(J \right)}(\hat{\rho}) > 0,
\,\,\,\,B -C \,A^{-1}\,C^{\dagger} > 0,
\label{4.30}
\end{eqnarray} 
where $A$, $B$, $C$ are taken from Eq.\,(\ref{4.29}). One can see that
some complication arises from the need to compute $A^{-1}$ in the new
condition. 

In the next Section, we analyse the case $J=\frac{1}{2}
\rightarrow J +\frac{1}{2}=1$ in some detail, as the first nontrivial
step going beyond the Schr\"{o}dinger-Robertson UP   
(\ref{3.7},\,\,\ref{4.28}). Before we turn to this task, however, a note on the 
non-generic case of singular $A$ seems to be in order.

\vskip0.2cm

\noindent
{\bf Remark}\,:  
Lemma 1 expresses the positive definiteness of a hermitian matrix $Q$
in the block form (\ref{2.10}) in terms of conditions on the lower
dimensional blocks. The block form itself is a description of $Q$
with respect to a given breakup of the underlying vector space on
which $Q$ acts, into two mutually orthogonal subspaces. Both $A$ 
and $B$ are hermitian. For the case of positive semidefinite $Q$,
there are two possibilities at the level of $A$, $B$, $C$. If $A^{-1}$
exists, then $Q\geq 0$ translates into $A > 0$,
$B-C\,A^{-1}\,C^{\dagger}\geq 0$. In case $A$ is singular, while of
course $Q \geq 0$ implies $A \geq 0$, the question is what other
condition on $B$, $C$ is implied. To answer this, we further separate the
subspace on which $A$ acts into two mutually orthogonal 
subspaces---one corresponding to the null subspace of $A$, and the other on 
which $A$ acts invertibly, say as $A_{1}$. Then in such a description,
the block form of $Q$ is initially refined to the form
\begin{eqnarray}
Q \simeq \left( \begin{array}{ccc}
0 & 0 & C_{2}^{\dagger} \\
0 & A_{1} & C_{1}^{\dagger} \\
C_{2} & C_{1} & B 
\end{array}
\right),
\end{eqnarray}
with the original $A$ and $C$ being respectively 
$\left(\begin{array}{cc} 0 & 0 \\ 0 & A_1\end{array} \right)$ 
and $(C_2,\,C_1)$. But now one sees easily that $Q \geq 0$ implies 
$C_2=0$, so as $A_1^{\,-1}$ exists, we have in this situation
\begin{eqnarray}
Q \geq 0 \,\,\,\Leftrightarrow \,\,\, A_{1} > 0,\,\,\, B-
C_{1}\,A_{1}^{-1}\, C_{1}^{\dagger} \geq 0.
\end{eqnarray}
This is the description of the nongeneric situation mentioned above.

\section{$SO(2,1)$ analysis of fourth order moments}
The first nontrivial step in the hierarchy of uncertainty relations
(\ref{4.27}), after the Schr\"{o}dinger-Robertson UP (\ref{3.7},\,\,\ref{4.28}),
occurs in going from $J =\frac{1}{2}$ to $J + \frac{1}{2}=1$. We study
this in some detail, especially as it brings into evidence the
equivalence of the irreducible representation $K^{(1)}(S)$ of
$Sp(2,\,R)$ and the defining representation of the three-dimensional
proper homogeneous Lorentz group $SO(2,1)$\,\cite{simon84}. Indeed $K^{(2)}(S)$,
$K^{(3)}(S)$, $\cdots$ are all true representations of $SO(2,1)$\,\cite{LCB}.

It is useful to introduce specific symbols for the operators
$\hat{T}_{\frac{1}{2}m}$, $\hat{T}_{1m}$ in the present context. We
write 
\begin{eqnarray}
(\hat{T}_{\frac{1}{2}m})&=&(\hat{\xi}_m) =\left( \begin{array}{c}
\hat{q} \\ \hat{p}
\end{array} \right),\,\,\,\,m=\frac{1}{2},\,-\frac{1}{2}\,; \nonumber \\
(\hat{T}_{1m})&=&(\hat{X}_{m})= \left( \begin{array}{c}
\hat{q}^2 \\ \frac{1}{2}\,\{\hat{q},\,\hat{p} \} \\ \hat{p}^2
\end{array} \right),\,\,\,\,m=1,\,0,\,-1\,;
\label{5.1}
\end{eqnarray}
so that one immediately recognises that $\hat{\xi}$ is a two-component
spinor, and $\hat{X}$ a three-component vector, with respect to
$SO(2,1)$ (see below). Their products can be computed using
Eq.\,(\ref{4.15}) or more directly by simple algebra\,:
\begin{equation}
\begin{array}{rclc}
\hat{\xi}_{m}\,\hat{\xi}_{m'} &=& \hat{X}_{m+m'} + i\,\frac{\hbar}{2}
(-1)^{m-\frac{1}{2}} \,\delta_{m+m',\,0}\,; & (a) \\
\hat{\xi}_{m}\,\hat{X}_{m'}&=& \hat{T}_{\frac{3}{2},\,m+m'} + i
\,\frac{\hbar}{2} (-1)^{m-\frac{1}{2}}\,\hat{\xi}_{m+m'}\,;
& (b) \\
\hat{X}_{m}\,\hat{X}_{m'} &=& \hat{T}_{2,\,m+m'} +
\frac{\hbar^2}{4}(-1)^m\, (1+m^2)\,\delta_{m+m',0} +
i\,\hbar\,(m-m')\,\hat{X}_{m+m'}. &(c)
\end{array}
\label{5.2}
\end{equation}
In ($5.2a$) the leading $J=1$ term is symmetric in $m$, $m^{\, \prime}$\,; while
the pure imaginary $J=0$ second term is antisymmetric. In ($5.2b$) it
is understood that $\hat{\xi}_{\pm \frac{3}{2}}=0$. In ($5.2c$) the
first two $J=2$ and $J=0$ terms are symmetric in $m$, $m^{\, \prime}$\,; while
the third $J=1$ term is antisymmetric. These features agree with the
pattern in Eq.\,(\ref{4.21}). 

For $J=\frac{1}{2}$ in Eq.\,(\ref{4.29}) we have 
\begin{eqnarray}
A=\tilde{\Omega}^{\left(\frac{1}{2},\,\frac{1}{2}
  \right)}(\hat{\rho}),\,\,\,\,
B=\tilde{\Omega}^{\left(1,\,1
  \right)}(\hat{\rho}),\,\,\,\,
C=\tilde{\Omega}^{\left( 1,\,\frac{1}{2}
  \right)}(\hat{\rho}),\,\,\,\,
\label{5.3}
\end{eqnarray}
with `dimensions' $2 \times 2$, $3 \times 3$, $3 \times 2$
respectively. (Throughout this Section, $A$, $B$, $C$ will have these
meanings). Their behaviours under $Sp(2,\,R)$ are 
\begin{eqnarray}
\hat{\rho}\rightarrow \hat{\rho}^{\, \prime}
=\overline{U}(S)\,\hat{\rho}\,\overline{U}(S)^{-1} &\Rightarrow & 
A \rightarrow S\,A\,S^{T},\,\,\,\,B \rightarrow K^{(1)}(S)\, B \,
K^{(1)}(S)^{T}, \nonumber \\
&& C \rightarrow K^{(1)}(S) \,C \,S^{T}.
\label{5.4}
\end{eqnarray}
Assuming $A^{-1}$ exists, we have
\begin{eqnarray}
A^{-1} \rightarrow (S^{-1})^{T}\, A^{-1}\, S^{-1},
\label{5.5}
\end{eqnarray}
and  consequently, 
\begin{eqnarray}
B- C \,A^{-1}\,C^{\dagger} &\rightarrow& K^{(1)}(S)\,(B - C\,
A^{-1}\,C^{\dagger})\,K^{(1)}(S)^{T},
\label{5.6}
\end{eqnarray}
which as expected is a congruence.

The matrix $K^{(1)}(S)$ is easily found. At the level of classical
variables\,:
\begin{eqnarray}
&&S=\left( \begin{array}{cc}
a & b\\ c&d
\end{array} \right) \in Sp(2,\,R)\,:\,\,\,\,
\left(\begin{array}{c} q \\ p \end{array} \right)
\rightarrow S\,\left(\begin{array}{c} q \\ p \end{array} \right)
\Rightarrow \nonumber \\
&&(X_{m}(q,\,p))= \left( \begin{array}{c} 
q^2 \\ qp \\ p^2
\end{array} \right) \rightarrow K^{(1)}(S)\,(X_{m}(q,p)), \nonumber \\
&&K^{(1)}(S)=\left( \begin{array}{ccc}
a^2 & 2ab & b^2 \\ ac & ad+bc & bd \\
c^2 & 2cd & d^2
\end{array} \right).
\label{5.7}
\end{eqnarray}
The link to $SO(2,1)$ can be seen in two (essentially equivalent)
ways, either through $A$ or through $(X_{m}(q,p))$. We now outline
both.

We introduce indices $\mu$, $\nu$, $\cdots$ going over values
$0,\,3,\,1$ (in that sequence) and a three-dimensional Lorentz metric
$g_{\mu \,\nu}={\rm diag}(+1,\,-1,\,-1)$. This metric and its inverse
$g^{\mu\,\nu}$ are used for lowering and raising Greek indices. The
defining representation of the proper homogeneous Lorentz group
$SO(2,1)$ is then\,:
\begin{eqnarray}
SO(2,1)=\{ \Lambda =({\Lambda^{\mu}}_{\nu})=3\times3 {\rm\,\,
  real\,\,matrix\,\,}&|&\,\,{\Lambda^{\mu}}_{\nu}\,\Lambda_{\mu \lambda}
\equiv g_{\mu
  \tau}\,{\Lambda^{\mu}}_{\nu}\,{\Lambda^{\tau}}_{\lambda}=g_{\nu
  \lambda},\nonumber \\
&&~~~{\rm det}\,\Lambda =+1,\,\,\,{\Lambda^{0}}_{0}\geq 1 \}.
\label{5.8}
\end{eqnarray} 
This is a three-parameter noncompact Lie group. Now expand
$A=\tilde{\Omega}^{\left(\frac{1}{2},\,\frac{1}{2}
  \right)}(\hat{\rho})$ in terms of Pauli matrices as follows\,:
\begin{eqnarray}
A=\tilde{\Omega}^{\left(\frac{1}{2},\,\frac{1}{2}
  \right)}(\hat{\rho}) = x^{\mu}\,\sigma_{\mu} -
\frac{\hbar}{2}\,\sigma_2 = 
\left(\begin{array}{cc}  
x^{0}+x^{3} & x^{1} \\
x^{1} & x^{0} - x^{3}
\end{array}\right) - \frac{\hbar}{2}\,\sigma_2 .
\label{5.9}
\end{eqnarray}
From Eqs.\,(\ref{3.6},\,\,\ref{4.28}) we have (indicating $\hat{\rho}$
dependences)\,:
\begin{eqnarray}
x^{0}(\hat{\rho})=\frac{1}{2}((\Delta q)^2 + (\Delta p)^2),\,\,\,
x^{3}(\hat{\rho})= \frac{1}{2}((\Delta q)^2 - (\Delta p)^2),\,\,\,
x^{1}(\hat{\rho})=\Delta(q,p)
\label{5.10}
\end{eqnarray}
Then the transformation rule for $A$ in Eq.\,(\ref{5.4}), combined with
$S\,\sigma_2 \,S^T = \sigma_2$, leads to a rule for
$x^{\mu}(\hat{\rho})$\,:
\begin{eqnarray}
&& \hat{\rho} \rightarrow
\overline{U}(S)\,\hat{\rho}\,\overline{U}(S)^{-1}\,\,\Rightarrow \,\,
A \rightarrow S\,A\,S^{T}\,\,\Rightarrow \,\, x^{\mu}(\hat{\rho})
\rightarrow {\Lambda^{\mu}}_{\nu}(S)\,x^{\nu}(\hat{\rho}), \nonumber \\
\Lambda(S)&=&\left( \begin{array}{ccc}
\frac{1}{2}(a^2 + b^2 + c^2 + d^2) &\frac{1}{2}(a^2 - b^2 + c^2 - d^2)
& ab + cd \\
\frac{1}{2}(a^2 + b^2 - c^2 - d^2) &\frac{1}{2}(a^2 - b^2 - c^2 + d^2)
& ab-cd \\
ac+bd & ac-bd & ad+bc  
\end{array} \right) \,\,\in SO(2,1)\,.
\label{5.11}
\end{eqnarray}
Thus $x^{\mu}(\hat{\rho})$ transforms as a Lorentz three-vector, and
the associated invariant is seen to be
\begin{eqnarray}
x^{\mu}(\hat{\rho})\,x_{\mu}(\hat{\rho})=g_{\mu \nu}
\,x^{\mu}(\hat{\rho})\,x^{\nu}(\hat{\rho}) =
(\Delta q)^2\,(\Delta p)^2 - (\Delta(q,p))^2 \geq \frac{\hbar^2}{4},
\label{5.12}
\end{eqnarray}
so the Schr\"{o}dinger-Robertson  UP implies the geometrical
statement that $x^{\mu}(\hat{\rho})$ is positive time-like.

The matrices $K^{(1)}(S)$ by which $\hat{X}_{m}$ transform under
$Sp(2,\,R)$ are related by a fixed similarity transform to the
$\Lambda(S)$ above. If in terms of classical variables we pass from
the components $X_{m}(q,p)$ in Eq.\,(\ref{5.7}) to a new set of
components $X^{\mu}(q,p)$ by
\begin{eqnarray}
(X^{\mu}(q,p))&=& \left( \begin{array}{c}
\frac{1}{2}(q^2 + p^2) \\ \frac{1}{2}(q^2 -p^2) \\qp
\end{array} \right) =
M\,\left( \begin{array}{c} 
q^2 \\ qp \\ p^2
\end{array} \right), \nonumber \\
X^{\mu}(q,p)&=& {M^{\mu}}_{m}\,X_{m}(q,p),\,\,\,\,X_{m}(q,p)=M^{-1}_{m
  \mu}\,X^{\mu}(q,p), \nonumber \\
M&=& ({M^{\mu}}_{m})=\left( 
\begin{array}{rrr}
\frac{1}{2} & ~0 & \frac{1}{2} \\
\frac{1}{2} & ~0 & - \frac{1}{2} \\
 0 & ~1 & 0 
\end{array}
\right),\,\,\,\, M^{-1}=(M^{-1}_{m \mu})=
\left( \begin{array}{rrr}
1 & 1 & ~0 \\
0 & 0 & ~1 \\
1 & -1 & ~0
\end{array} \right),
\label{5.13}
\end{eqnarray}
then in place of Eq.\,(\ref{5.7}) we have 
\begin{eqnarray}
\left( \begin{array}{c}q \\p \end{array} \right)
\rightarrow S\,\left( \begin{array}{c}q \\p \end{array} \right)
\Rightarrow X^{\mu}(q,p)&\rightarrow &{M^{\mu}}_{m} \, K^{(1)}_{mm'}(S)\,
M^{-1}_{m' \nu}\,X^{\nu}(q,p) \nonumber \\
&=& {\Lambda^{\mu}}_{\nu}(S)\,X^{\nu}(q,p), \nonumber \\
K^{(1)}(S)&=& M^{-1}\,\Lambda(S)\,M.
\label{5.14}
\end{eqnarray}
At the operator level we have 
\begin{eqnarray}
&&\hat{X}^{0}=\frac{1}{2}(\hat{q}^2 + \hat{p}^2),\,\,\,\hat{X}^{3}=
\frac{1}{2}(\hat{q}^2 -
\hat{p}^2),\,\,\,\hat{X}^1=\frac{1}{2}\{\hat{q},\, \hat{p} \},\nonumber \\
&& \hat{X}^{\mu}= {M^{\mu}}_{m}\,\hat{X}_{m},
\label{5.15}
\end{eqnarray}
and, as consequence of Eq.\,(\ref{4.13})), the twin equivalent transformation laws\,:
\begin{eqnarray}
S \in Sp(2,\,R)\,:~~&&
\overline{U}(S)^{-1}\,\hat{X}_{m}\,\overline{U}(S)=K^{(1)}_{mm'}(S)
\,\hat{X}_{m'} , \nonumber \\
&&\overline{U}(S)^{-1}\,\hat{X}^{\mu}\,\overline{U}(S)={\Lambda^{\mu}}_{\nu}(S)
\,\hat{X}^{\nu}. 
\label{5.16}
\end{eqnarray}
The upshot is that the matrices $K^{(1)}(S)$ are just the `ordinary'
homogeneous Lorentz transformation matrices $\Lambda(S)$ in a `tilted'
basis. The metric preserved by them is easily found though
unfamiliar\,:
\begin{eqnarray}
&&K^{(1)}(S)\,g_K \,K^{(1)}(S)^T = g_K , \nonumber \\
&& g_{K}=M^{-1}\,g\, (M^{-1})^T = 
\left( \begin{array}{ccc}
0 & 0 & 2 \\ 0 &-1 & 0 \\ 2 & 0 &0
\end{array} \right).
\label{5.17}
\end{eqnarray}
This enables us to use the nomenclature and geometrical features of
three-dimensional Minkowski space even while working with operators 
$\hat{X}_{m}$ and transformation matrices $K^{(1)}(S)$.

Now we proceed to analyse the three matrices $A$, $B$, $C$ and the
combination $B- C\,A^{-1}\,C^{\dagger}$. (We have already parametrised
$A$ in Eqs.\,(\ref{5.9},\,\,\ref{5.10})). Using Eqs.\,(\ref{5.2}), their matrix
elements are 
\begin{eqnarray}
A_{mm'}&=& \langle\hat{\xi}_{m}\,\hat{\xi}_{m'} \rangle -
\langle\hat{\xi}_{m} \rangle \,\langle \hat{\xi}_{m'} \rangle
\nonumber \\
&=& \langle \hat{X}_{m+m'} \rangle -
\langle\hat{\xi}_{m} \rangle \,\langle \hat{\xi}_{m'} \rangle
+i\,\frac{\hbar}{2}\,(-1)^{m-\frac{1}{2}}\,\delta_{m,\,-m'} \nonumber \\
&=&(x^{\mu}\,\sigma_{\mu})_{mm'} +i
\,\frac{\hbar}{2}\,(-1)^{m-\frac{1}{2}}\,\delta_{m,\,-m'}\,; \nonumber
\\
B_{mm'}&=& \langle\hat{X}_{m}\,\hat{X}_{m'} \rangle -
\langle\hat{X}_{m} \rangle \,\langle \hat{X}_{m'} \rangle
\nonumber \\
&=& \langle \hat{T}_{2,m+m'} \rangle +
\frac{\hbar^2}{4}\,(-1)^{m}\,\delta_{m,-m'}-
\langle\hat{X}_{m} \rangle \,\langle \hat{X}_{m'} \rangle + i\,\hbar
\,(m-m')\,\langle \hat{X}_{m+m'} \rangle ; \nonumber \\
C_{mm'}&=& \langle\hat{X}_{m}\,\hat{\xi}_{m'} \rangle -
\langle\hat{X}_{m} \rangle \,\langle \hat{\xi}_{m'} \rangle
\nonumber \\
&=& \langle \hat{T}_{\frac{3}{2},m+m'} \rangle -
\langle\hat{X}_{m} \rangle \,\langle \hat{\xi}_{m'} \rangle
- i\,\frac{\hbar}{2}\,(-1)^{m'-\frac{1}{2}}\,\langle \hat{\xi}_{m+m'} \rangle.
\label{5.18}
\end{eqnarray}
In each of these expressions, the possible values for $m$, $m^{\, \prime}$ are
evident from the context. We now note an important fact in respect of the final
forms of all three expressions\.: apart from explicit appearances of $i$
in the last terms, {\em all other quantities are real}. This
allows us to easily separate each of $A$, $B$, $C$ into real and imaginary
parts, which in the cases of $A$ and $B$ are respectively symmetric and
antisymmetric in $m$ and $m^{\, \prime}$. [\,This is already seen in Eq.\,(\ref{5.9})
for $A$\,]. We write these as follows\,:
\begin{equation}
\begin{array}{rclc}
A &=& A_{1}+ i\,A_{2},   &  \\
(A_{1})_{mm'}&=& \langle\hat{X}_{m+m'} \rangle -
\langle\hat{\xi}_{m} \rangle \,\langle \hat{\xi}_{m'} \rangle
=(x^{\mu}\,\sigma_{\mu})_{mm'}\,,
 &  \\
(A_{2})_{mm'}&=& \,\frac{\hbar}{2}\,(-1)^{m-\frac{1}{2}}\,\delta_{m,\,-m'}; 
  & ~~(a) \\
B&=&B_{1} + i\,B_{2}, & \\
(B_{1})_{mm'}&=& \langle \hat{T}_{2,m+m'} \rangle +
\frac{\hbar^2}{4}\,(-1)^{m}\,\delta_{m,-m'}-
\langle\hat{X}_{m} \rangle \,\langle \hat{X}_{m'} \rangle,  & \\
(B_{2})_{mm'} & =& \,\hbar
\,(m-m')\,\langle \hat{X}_{m+m'} \rangle ; & ~~(b) \\
C&=&C_{1}+ i\,C_{2};  &  \\
(C_{1})_{mm'}&=& \langle \hat{T}_{\frac{3}{2},m+m'} \rangle -
\langle\hat{X}_{m} \rangle \,\langle \hat{\xi}_{m'} \rangle,  &  \\
(C_{2})_{mm'}&=& - \,\frac{\hbar}{2}\,(-1)^{m'-\frac{1}{2}}\,\langle
\hat{\xi}_{m+m'} \rangle\,; &  \\
C^{\dagger}&=& C_{1}^{T} - i\,C_{2}^{T}\,. & ~~(c)
\end{array}
\label{5.19}
\end{equation}
To deal similarly with $B - C\,A^{-1} \,C^{\dagger}$, we need an expression
for $A^{-1}$. We will assume the generic situation in which $A$ is
nonsingular,
\begin{eqnarray}
{\rm det}\,A \equiv \kappa^{-1}=x^{\mu}\,x_{\mu} - \frac{\hbar^2}{4} > 0,
\label{5.20}
\end{eqnarray}
so that 
\begin{eqnarray}
A^{-1}&=& \kappa (x^0 - x^3 \sigma_3 -x^1 \sigma_1 + \frac{\hbar}{2}\,\sigma_2)
\nonumber \\
&=& \kappa(\tilde{x}^{\mu}\,\sigma_{\mu} + \frac{\hbar}{2}\,\sigma_2),
\nonumber \\
\tilde{x}^{\mu}&=& (x^0,\,-x^3,\,-x^1).
\label{5.21}
\end{eqnarray}
The transformation law for $A^{-1}$ under $S \in Sp(2,\,R)$ given in
Eq.\,(\ref{5.5}) is different from (though equivalent to) the law for
$A$. 
Thus, while  the $\tilde{x}^{\mu}$ do follow a 
definite (i.e., well defined tensorial\,\cite{simon84}) 
transformation law, there are some differences (in signs) compared to the law followed by  
$x^{\mu}$.  
 Clearly the two terms in Eq.\,(\ref{5.21}) are, as they
stand, the real symmetric and the pure imaginary antisymmetric parts
of $A^{-1}$. We can now handle $B- C\, A^{-1}\,C^{\dagger}$ in the
same manner as above\,:
\begin{eqnarray}
B - C\,A^{-1}\,C^{\dagger} &=& B_{1}+ i\,B_{2}- \kappa (C_{1}+
i\,C_{2})\,(\tilde{x}\cdot \sigma
+\frac{\hbar}{2}\,\sigma_2)\,(C_{1}^{T}-i\,C_{2}^{T}) \nonumber \\
&=&V^{({\rm eff})}+ \frac{i}{2} \,\omega^{({\rm eff})}, \nonumber \\
V^{({\rm eff})}&=& B_{1} - \kappa(C_{1}\,\tilde{x}\cdot\sigma\,C_{1}^{T}
+C_{2}\,\tilde{x}\cdot\sigma C_{2}^{T} + i \,\frac{\hbar}{2}
C_{2}\,\sigma_{2}\,C_{1}^{T} -i\,\frac{\hbar}{2}\,
C_{1}\,\sigma_2\,C_{2}^{T}), \nonumber \\
\frac{1}{2}\,\omega^{({\rm eff})}&=& B_{2} - 
\kappa(C_{2}\,\tilde{x}\cdot\sigma\,C_{1}^{T}
- C_{1}\,\tilde{x}\cdot\sigma C_{2}^{T} - i \,\frac{\hbar}{2}
C_{1}\,\sigma_{2}\,C_{1}^{T} -i\,\frac{\hbar}{2}\,
C_{2}\,\sigma_2\,C_{2}^{T}). ~~
\label{5.22}
\end{eqnarray}
This decomposition is in the spirit and notation of Eq.\,(\ref{2.18}) of
the general framework. However, 
$V^{({\rm eff})}$ and  $\omega^{({\rm eff})}$ do not correspond any longer to 
expectation values of simple anticommutators and commutators among relevant operators, 
as was the case in Eqs.\,(\ref{2.18},\,\ref{4.18},\,\ref{4.20}). 

Both $V^{({\rm eff})}$ and $\omega^{({\rm eff})}$ are
real three-dimensional matrices with elements $V^{({\rm eff})}_{m m'}$,
$\omega^{({\rm eff})}_{m m'}$, where $m,\,m^{\, \prime} = 1,\,0,\,-1$\,; and they are
respectively symmetric and antisymmetric. It does not seem possible to
simplify the expressions (\ref{5.22}) to any significant extent, as they
are already expressed in terms of the independent real expectation
values $\langle \hat{\xi}_m \rangle$, $\langle \hat{X}_{m} \rangle$,
$\langle \hat{T}_{\frac{3}{2}, m} \rangle$, $\langle \hat{T}_{2,m}
\rangle$ which are the moments of the Wigner distribution $W(q,\, p)$
of orders up to and including the fourth. Under action by $S\in
Sp(2,\,R)$ we have from Eq.\,(\ref{5.6})\,:
\begin{eqnarray}
\hat{\rho}\rightarrow \overline{U}(S)\,\hat{\rho}\, \overline{U}(S)^{-1}\,\, 
&\Rightarrow& \,\,V^{\rm (eff)} \rightarrow K^{(1)}(S)\,V^{\rm
  (eff)}\,K^{(1)}(S)^{T}, \nonumber \\
&& \,\,\omega^{\rm (eff)} \rightarrow K^{(1)}(S)\,
\omega^{\rm (eff)}\,K^{(1)}(S)^{T}.
\label{5.23}
\end{eqnarray} 
The added uncertainty relation up to the fourth order going beyond the
Schr\"{o}dinger-Robertson UP (\ref{3.7},\,\,\ref{4.28}), 
reads [\,in the
generic case ${\rm det}\,A > 0$\,]\,:
\begin{eqnarray}
V^{\rm (eff)} + \frac{i}{2}\,\omega^{\rm (eff)} \geq 0,
\label{5.24}
\end{eqnarray}
 which is an $SO(2,1)$ covariant statement 
by virtue of Eq.\,(\ref{5.23}). 

For further analysis it is rather awkward to work with $SO(2,1)$
matrices and Lorentz metric in the form $K^{(1)}(S)$, $g_{K}$,
therefore we pass to the `standard' forms via the matrices $M$,
$M^{-1}$ in Eq.\,(\ref{5.13})\,:
\begin{eqnarray}
V^{\rm (eff)\,\mu \nu} ={M^{\mu}}_{m}\,{M^{\nu}}_{m'}\,V^{\rm (eff)}_{m
  m'}, \nonumber \\
\omega^{\rm (eff)\,\mu \nu} = {M^{\mu}}_{m}\,{M^{\nu}}_{m'}\,\omega^{\rm (eff)}_{m
  m'}\,,
\label{5.25}
\end{eqnarray}
which are congruences. 
Then the $Sp(2,\,R)$ or $SO(2,\,1)$ actions (\ref{5.23}) appear as\,:
\begin{eqnarray}
&&V^{\rm (eff)\,\mu \nu} \rightarrow
{\Lambda^{\mu}}_{\mu'}(S)\,{\Lambda^{\nu}}_{\nu'}(S)\, V^{\rm (eff)\, \mu'
\nu'}, \nonumber \\
&&\omega^{\rm (eff)\,\mu \nu} \rightarrow
{\Lambda^{\mu}}_{\mu'}(S)\,{\Lambda^{\nu}}_{\nu'}(S)\, \omega^{\rm (eff)\, \mu'
\nu'}, 
\label{5.26}
\end{eqnarray}
and the condition (\ref{5.24}) becomes\,:
\begin{eqnarray}
(V^{\rm (eff) \,\mu \nu}) + \frac{i}{2}\,(\omega^{\rm (eff)\, \mu
    \nu}) \geq 0.
\label{5.27}
\end{eqnarray}
While $V^{\rm (eff)\,\mu \nu}$ transforms as a symmetric second rank
$SO(2,1)$ tensor, $\omega^{\rm (eff)\,\mu \nu}$ is an antisymmetric
second rank tensor, which by the use of the Levi Civita invariant
tensor is the same as a three vector. Thus we can write, with
$\epsilon^{031}=\epsilon_{031} =+1$,
\begin{eqnarray}
\omega^{\rm (eff)\,\mu \nu} &=& \epsilon^{\mu \nu \lambda}\, a_{\lambda},
\nonumber \\
(\omega^{\rm (eff)\,\mu \nu})&=& \left( \begin{array}{ccc}
0 & a_1 & -a_3 \\
-a_1 & 0 & a_0 \\
a_3 & -a_0 & 0
\end{array} \right),
\label{5.28}
\end{eqnarray}
with transformation law
\begin{eqnarray}
a^{\mu} \rightarrow {\Lambda^{\mu}}_{\nu}(S)\, a^{\nu}.
\label{5.29}
\end{eqnarray}
Of course, $V^{\rm (eff) \,\mu \nu}$ itself is made up of 
two irreducible parts\,: the symmetric second rank `trace-free' part 
belonging to the $SO(2,\,1)$ representation $K^{(2)}(S)$, and the $SO(2,\,1)$ 
 invariant trace which is a scalar. 
 
We now appeal to a remarkable result\,\cite{RSSCVS}, which is 
similar in  spirit to the Williamson theorem
for $Sp(2n,\,R)$ quoted in Section 3. It states that if $V^{\rm
  (eff)\,\mu \nu}$ transforming as in Eq.\,(\ref{5.26}) is positive
definite, it is possible to bring it to a diagonal form by a suitable
choice of $\Lambda \in SO(2,1)$\,; however in general the resulting
diagonal values are not the eigenvalues of the initial matrix. This
diagonal form may be called the `SCS normal form' of $V^{\rm (eff)}$,
which in the generic case is unique. Passing to this normal form of
$V^{\rm (eff)}$, and transforming $\omega^{\rm (eff)}$ as well by the
same (generically unique) $\Lambda \in SO(2,1)$, these matrices appear
as 
\begin{eqnarray}
V^{\rm (eff)} \rightarrow \left(\begin{array}{ccc}
v^{00} & 0 & 0 \\ 0 & v^{33} & 0 \\ 0 & 0 & v^{11}
\end{array} \right),\,\,\,
\omega^{\rm (eff)}\rightarrow \left( \begin{array}{ccc}
0 & -b^1 & b^3 \\ b^1 & 0 & b^0 \\-b^3 & -b^0 &0
\end{array} \right),
\label{5.30}
\end{eqnarray}
with all the quantities $v^{00}$, $v^{33}$, $v^{11}$, $b^0$, $b^3$,
$b^1$ being real $SO(2,1)$ (and $Sp(2,\,R)$) invariants. The
uncertainty relation (\ref{5.27}) expressed in terms of these invariants,
and in its maximally simplified form thanks to the SCS theorem, is
\begin{eqnarray}
\left(\begin{array}{ccc}
v^{00} & 0 & 0 \\ 0 & v^{33} & 0 \\ 0 & 0 & v^{11}
\end{array} \right) +\frac{i}{2}\, 
\left( \begin{array}{ccc}
0 & -b^1 & b^3 \\ b^1 & 0 & b^0 \\-b^3 & -b^0 &0
\end{array} \right)\geq 0.
\label{5.31}
\end{eqnarray}


As an (admittedly elementary) example of the discussion of this
Section, we consider the Fock states $|n \rangle$, $n \geq 0$. The
$(\hat{q},\,\hat{p})$ --- $(\hat{a},\,\hat{a}^{\dagger})$ relations are 
\begin{align}
\hat{a} = (\hat{q} + i \hat{p})/\sqrt{2 \hbar}, ~~~ \hat{a}^{\dagger} = (\hat{q} - i \hat{p})/\sqrt{2 \hbar}, 
\label{5.32}
\end{align}
so both $\hat{q}$ and $\hat{p}$ have dimensions $\hbar^{{{{1}/{2}}}}$. In the
Fock states $|n\rangle$, by parity arguments we have 
\begin{align} 
\langle n | \hat{\xi}_m |n\rangle = \langle n | \hat{T}_{3/2,m}| n
  \rangle =0.
\label{5.33}
\end{align}
For $\hat{X}_{m}, \, \hat{T}_{2,m}$ explicit calculations give\,:
\begin{align}
\langle n | \hat{X}_m | n \rangle &= \hbar (n+{{\frac{1}{2}}}) (1,\,0,\,1), 
~~ m = 1,\,0,\,-1;\nonumber\\
\langle n | \hat{T}_{2,m} | n \rangle &= \frac{1}{2} \hbar^2 (n^2 + n+
  {{\frac{1}{2}}}) (3,\,0,\,1,\,0,\,3), ~~ m = 2,\,1,\,0,\,-1,\,-2.
\label{5.34}
\end{align}
Then the matrices $A,\, B,\, C$ of Eq.\,\eqref{5.3} follow easily\,:
\begin{align}
\begin{array}{rclc}
\left(A_{mm^{\,'}}\right) &=& \hbar (n+{{\frac{1}{2}}}) 1\!\!1 
       - \frac{\hbar}{2}\, \sigma_2, & \\
x^0 &=& \hbar (n+{{\frac{1}{2}}}),  ~~ x^3 = x^1 =0; & (a)\\
\left(B_{mm^{\,'}}\right) &=& \frac{\hbar^2}{2} (n^2 + n +1
) \left( \begin{array}{rrr}1& ~0&-1 \\ 0&1&0 \\ -1&0&1 \end{array} \right) + i \hbar^2
(n + {{\frac{1}{2}}}) \left( \begin{array}{rrr} 0&1&~0 \\ 
              -1&0 &1 \\ 0 & -1 & 0\end{array} \right); & ~~(b) \\
\left( C_{mm^{\,'}}\right) &=& 0. & (c) 
\end{array}
\label{5.35}
\end{align}
Therefore the combination $B - CA^{-1} C^{\dagger} = B$, and from
Eq.\,\eqref{5.22}, 
\begin{align} 
\left(V^{(\rm eff)}_{mm^{\,'}} \right) &= \frac{\hbar^2}{2} (n^2 + n+
1) \left( \begin{array}{rrr} 1& ~0&-1 \\ 0&1&0 \\ -1&0&1\end{array} \right), \nonumber\\
\frac{1}{2}\left(\omega^{(\rm eff)}_{mm^{\,'}} \right) &=
\hbar^2 (n+{{\frac{1}{2}}}) 
   \left( \begin{array}{rrr} 0&1& ~0 \\ -1&0 &1 \\ 0 & -1 & 0\end{array} \right). 
\label{5.36}
\end{align}
Transforming to the standard $SO(2,1)$ tensor components by the
congruence transformation\,\eqref{5.25} we find\,:
\begin{align} 
\left(V^{(\rm eff)\,\mu\nu} \right) &= \frac{\hbar^2}{2} (n^2 + n+
1) \left( \begin{array}{ccc} 0&~0~&0 \\ 0&1&0 \\ 0&0&1\end{array} \right), \nonumber\\
\frac{1}{2}\left(\omega^{(\rm eff)\,\mu\nu} \right) &=
\hbar^2 (n+{{\frac{1}{2}}}) \left( \begin{array}{rrr} 0&0&~0 \\ 0&0 &1 \\ 0 & -1 & 0\end{array} \right). 
\label{5.37}
\end{align}
As expected, both these matrices are invariant under the $SO(2)$ subgroup of 
$SO(2,\,1)$, as the Fock states are eigenstates of the phase space rotation generator 
 $\hat{a}^\dagger\hat{a}$. 
   
We see that $\left(V^{(\rm eff)\,\mu\nu} \right)$ is already in the
$SCS$ normal form, and as the eigenvalues of $\left(V^{(\rm
  eff)\,\mu\nu} \right) + \frac{\,i\,}{2}\, \left(\omega^{(\rm eff)\,\mu\nu} \right)$
are $\,0,\, \frac{\hbar^2}{2}(n^2+n+1) \pm \hbar^2 (n+ {{\frac{1}{2}}})$, 
\,i.e., $\,0,\,
\frac{\hbar^2}{2}(n+1)(n+2),\, \frac{\hbar^2}{2} n(n-1)$, the
uncertainty relation\,\eqref{5.27} is clearly respected; indeed it is saturated!

\section{Lorentz geometry and the Schr\"{o}dinger-Robertson UP} 
The original Schr\"{o}dinger-Robertson UP  has a very interesting character 
when viewed in the Wigner distribution language, bringing out the role of the group 
$SO(2,\,1)$ in a rather striking manner. This seems worth exploring in some detail. 

For a given state $\hat{\rho}$ with Wigner distribution $W(q,\,p)$, the means are  
\begin{align}
\overline{q} = \int\int dqdp\,q\,W(q,\,p)\,, ~~~
\overline{p} = \int\int dqdp\,p\,W(q,\,p)\,.
\label{6.1}
\end{align}
Referring to Eq.\,\eqref{5.13}, at each point $(q,\,p)$ in the 
phase plane we define the $SO(2,\,1)$ three-vector (a displaced form
of $(X^{\mu}(q,p))$ in Eq.\,\eqref{5.13})
\begin{align}
(\,X^\mu(q,\,p)\,) = \left(\begin{array}{c}
\frac{1}{2}\,[\,(q - \overline{q})^2 + (p - \overline{p})^2\,]\\
 \frac{1}{2}\,[\,(q - \overline{q})^2 - (p - \overline{p})^2\,]\\
(q - \overline{q})(p - \overline{p})
\end{array}
\right),
\label{6.2}
\end{align}
which  (\,except at $q = \overline{q},\,p=\overline{p}$\,) 
is pointwise positive light-like. The elements of the variance
matrix $V$ in Eqs.\,($3.6, \, 4.29$) are obtained by `averaging' this 
three-vector over the phase plane with the quasiprobability 
$W(q,p)$ as `weight' function, resulting in the three-vector
$x^{\mu}(\hat{\rho})$ of Eq.\,\eqref{5.10}\,:
\begin{align}
(x^{\mu}(\hat{\rho})) = 
\left( \begin{array}{c}
\frac{1}{2}\, [\,(\triangle q )^2+(\triangle p)^2\,]\\
\frac{1}{2}\, [\,(\triangle q )^2-(\triangle p)^2\,]\\
\triangle (q,p)
\end{array}
\right) = \int \int dqdp\, W(q,p)\, (\,X^\mu(q,\,p)\,).
\label{6.3}
\end{align}
Given that $W(q,p)$ can in principle be negative over certain regions
of the phase space, this `averaging' could have led to a result which
need not be either time-like or light-like positive. However the
Schr\"{o}dinger-Robertson UP assures us that in fact the result has to be
a time-like positive three-vector, thus implying a subtle limit on the
extent to which $W(q,p)$ could become negative.
 In fact it specifies that the three-vector obtained as a result of the `averaging' 
 must be within or on the positive time-like (solid) hyperboloid $\sum_{\mu} x^{\mu}(\hat{\rho}) x_{\mu}(\hat{\rho}) \geq \hbar^2/4$ corresponding 
 to `squared mass' $\hbar^2/4$ presented in Eq.\,\eqref{5.12}. On the other hand, while pointwise nonnegativity of $W(q,p)$ will 
certainly ensure that the averaging in Eq.\,\eqref{6.3} takes $\left( x^{\mu}(\hat{\rho})\right)$ inside the time-like
positive cone, it will not itself ensure that it is taken all the way
inside the said hyperboloid. To ensure the latter, $W(q,p)$ should have a threshold effective spread.
Thus, pointwise nonnegativity is neither a necessary nor sufficient requirement to ensure `Wigner quality' 
on $W(q,p)$ as is known from other considerations. 


The argument above has been presented in such a way that the
interpretation in terms of Lorentz geometry in $2+1$ dimensions is
obvious. However, comparing Eqs.\,\eqref{5.7} and \eqref{5.13}, we
see that it could be expressed equally well as follows. At each point
$(q,p)$ in the phase plane we define a $2\times 2$ real symmetric matrix 
\begin{align}
V(q,p) = \left( \begin{array}{c}  q -\overline{q}
  \\ p-\overline{p} \end{array} \right) \, 
 \begin{array}{c} 
\left( \begin{array}{cc} q-\overline{q} & p-\overline{p} \end{array}
\right) \\ ~~~~
\end{array} .
\label{6.4}
\end{align}
Pointwise (except at $q = \overline{q},\, p = \overline{p}$ ) this is
proportional to a one-dimensional projection matrix, and in particular
it has vanishing determinant. After `averaging' with $W(q,p)$ as
weight function, however, we obtain the $2 \times 2$ variance matrix
$V$ in Eq.\,\eqref{4.28}\,:
\begin{align}
V = \int \int dp\,dq \, W(q,p) V(q,p) = \left( \begin{array}{cc} 
(\triangle q)^2 & \triangle(q,p) \\
\triangle(q,p) & (\triangle p)^2
\end{array}
\right),
\label{6.5}
\end{align}
and now the Schr\"{o}dinger-Robertson UP shows that $V$ is non-singular
and has determinant bounded below by the `squared mass' $\hbar^2/4$. 

In this form, just like the Schr\"{o}dinger-Robertson UP, this geometrical
picture based on the Wigner distribution language generalises in both
directions---second order moments for a multi mode system, and higher
order moments for a single mode system. As an example of the former,
consider a two-mode system for simplicity. The classical phase space
variables are $\xi_a$ and the hermitian quantum operators obeying
Eq.\,\eqref{3.1} are $\hat{\xi}_a$, for $a=1,\cdots,4$. Given
 a two-mode state $\hat{\rho}$, we pass to its Wigner
distribution $W(\xi)$ (something we did not do in Section III) and
compute the means
\begin{align}
\langle \hat{\xi}_a \rangle = {\rm Tr}(\hat{\rho}\,\hat{\xi}_a) = \int
d^4 \xi \, \xi_a W(\xi) = \overline{\xi}_a,\, a = 1,\cdots,4.  
\label{6.6}
\end{align}  
Then, generalising Eq.\,\eqref{6.4} above, at each point $\xi$
 in the 4-dimensional phase space we define a real symmetric $4\times
 4 $ matrix 
\begin{align}
V(\xi) &= (V_{ab}(\xi)) = ((\xi_a - \overline{\xi}_a)(\xi_b -
\overline{\xi}_b)) = x(\xi) x(\xi)^T,\nonumber\\
x_a(\xi) &= \xi_a - \overline{\xi}_a.
\label{6.7}
\end{align}
At each point $\xi$ (except at $\xi = \overline{\xi}$) we have here a
real symmetric positive semidefinite matrix $V(\xi)$ which is
essentially a one-dimensional projection matrix\,: the eigenvalues 
of $V(\xi)$ are $x(\xi)^T x(\xi),0,0,0$. The variance matrix $V$ for
the state $\hat{\rho}$ is then obtained by `averaging' $V(\xi)$ using
the real normalised quasiprobability $W(\xi)$\,:
\begin{align}
V = \int d^4 \xi \, W(\xi) V(\xi).
\label{6.8}
\end{align}
Since in general $W(\xi)$ can assume negative values at some places in
phase space, it may appear at first sight that some of the properties of
$V(\xi)$ described above may be lost by the `averaging' process
leading to $V$. However the UP\,\eqref{3.5} guarantees that this will
not happen; indeed by Lemma 2, Section II, in Eq.\,\eqref{2.13}, $V$ is seen
to be positive definite. Quantitatively we have the following
situation\,: Williamson's theorem assures us that under the congruence
transformation by a suitable $S_0 \in Sp(4, {\cal R})$, $V $ is taken
to a diagonal form\,:
\begin{align}
V_0 = S_0 V S_0^T = diag(\kappa_1,\kappa_1,\kappa_2,\kappa_2),
~~\kappa_{1,2} >0.
\label{6.9}
\end{align}
The congruence transformation becomes a similarity transformation on
$V \beta^{-1} $\,\cite{dutta94}, since\,:
\begin{align}
S \in Sp(4,{\cal R}) \,: V^{\,'} = SVS^T \leftrightarrow V^{\,'}
\beta^{-1} = S V \beta^{-1} S^{-1}.
\label{6.10}
\end{align}
Applying this to the transition $V \to V_0$ we see that as 
\begin{align}
V_0 \beta^{-1} = -i \left( \begin{array}{cc} 
\kappa_1 \sigma_2 &0 \\
0& \kappa_2 \sigma_2
\end{array} \right) ,
\label{6.11}
\end{align}
the eigenvalues of $i V \beta^{-1}$ are $\pm \kappa_1, \, \pm
\kappa_2 $; and the UP\,\eqref{3.5} ensures that $\kappa_{1,2} \geq
\hbar/2$. The $\kappa$'s themselves are determined, upto an
interchange, by the $Sp(4,{\cal R})$ invariant traces
\begin{align}
{\rm Tr}(V\beta^{-1})^2 = -2(\kappa_1^2 + \kappa_2^2),\nonumber\\
{\rm Tr}(V\beta^{-1})^4 = 2(\kappa_1^4 + \kappa_2^4).
\label{6.12}
\end{align}
The manner in which the geometrical picture, and the constraint on the
extent to which $W(\xi)$ can become negative, both generalise in going to
the multi mode situation is now clear. 

A qualitatively similar situation (even if algebraically more
involved) obtains when we generalise in the other direction---to
higher order moments for a single mode system, and their uncertainty
relations handled in the Wigner distribution language. Limiting
ourselves to the moments upto fourth order, we are concerned in the
notation of Eq.\,\eqref{4.23} with the uncertainty relation 
\begin{align}
\tilde{\Omega}^{(1)}(\hat{\rho}) \geq 0
\label{6.13}
\end{align}
contained in the hierarchy\,\eqref{4.27}, and its rendering in the
Wigner distribution language. Combining the notations of Sections II
and V, we have a set of five real phase space functions $A_{a}(q,p)$,
$a=1,2,\cdots,5$, and their hermitian operator counterparts in the
Weyl sense\,:
\begin{align}
(A_a(q,p)) &= (q,\,p,\,q^2,\,qp,\,p^2)^T;\nonumber\\
(\hat{A}_a) &= ((A_a(q,p))_W) =
  (\hat{q},\,\hat{p},\,\hat{q}^2,\,\frac{1}{2} \{\hat{q},\hat{p}\},\,\hat{p}^2)^T,
\label{6.14}
\end{align}
a listing of the components $\hat{\xi}_m$, $\hat{X}_m$. In a given
state $\hat{\rho}$ with Wigner distribution $W(q,p)$ we have the means 
\begin{align}
\langle \hat{A}_a \rangle = {\rm Tr}(\hat{\rho} \hat{A}_a) = \int \int
dp\,dq \, W(q,p) A_a(q,p) = \overline{A}_a.
\label{6.15}
\end{align}
To calculate the elements of $\tilde{\Omega}^{(1)}(\hat{\rho})$ we
need to deal with the products $\hat{A}_a \hat{A}_b $. For these,
using Eq.\,\eqref{5.2} we find\,:
\begin{align}
\hat{A}_a \hat{A}_b &= (A_a(q,p)A_b(q,p))_W +
(C_{ab}(q,p))_W,\nonumber\\
(C_{ab}(q,p)) &= \left( \begin{array}{ccccc} 
0 & \frac{i\hbar}{2} &~ 0 ~& \frac{i\hbar q }{2} & \frac{i\hbar p }{2}\\
-\frac{i\hbar}{2} & 0 &~ -\frac{i\hbar q}{2} ~& -\frac{i\hbar p}{2} & 0
\\
0 & \frac{i\hbar q }{2} &~ 0~& i\hbar q^2 &~ -\frac{\hbar^2}{2} + 2i \hbar
q p ~\\
-\frac{i\hbar q}{2} & \frac{i\hbar p}{2} &~ -i\hbar q^2 ~&
\frac{\hbar^2}{4} & i\hbar p^2 \\
-\frac{i\hbar p}{2} & 0 &~~-\frac{\hbar^2}{2} - 2i \hbar
q p~ ~& -i \hbar p^2 & 0                                                                  
\end{array}
\right).
\label{6.16}
\end{align} 
(We note that the real symmetric part of the matrix $C(q,p)$ is
$-\frac{\hbar^2}{4} g_K$ in the lower $3 \times 3$ block, where $g_K$
is the tilted form of the $(2+1)$ Lorentz metric in
Eq.\,\eqref{5.17}). 
With these ingredients and referring to the general
structure\,\eqref{2.16} we have the expression for
$\tilde{\Omega}^{(1)}(\hat{\rho})$ in the Wigner distribution
language\,:
\begin{align}
\tilde{\Omega}^{(1)}(\hat{\rho}) &= 
\left(\tilde{\Omega}^{(1)}_{ab}(\hat{\rho})\right) = \left({\rm Tr}(\hat{\rho}(\hat{A}_a - \langle \hat{A}_a \rangle)(\hat{A}_b - \langle \hat{A}_b \rangle)) \right) \nonumber\\
&= \left({\rm Tr}(\hat{\rho}\,(\,(A_a(q,p)A_b(q,p))_W -
\overline{A}_a\overline{A}_b  + (C_{ab}(q,p))_W)\,)\right) \nonumber\\
&= \int \int dp dq \,W(q,p)\left( x(q,p) x(q,p)^T + (C_{ab}(q,p))\right),\nonumber\\
x(q,p)^T&=(q-\overline{q},\,p-\overline{p},\,q^2-\overline{q^2},\,
qp-\overline{qp},\,p^2-\overline{p^2} ).
\label{6.17}
\end{align} 
At each point $(q,p)$ in the phase plane, we have essentially a one-dimensional 
projector $x(q,p) x(q,p)^T$, together with a five-dimensional hermitian 
matrix $(C_{ab}(q,p))$ with elements involving
$\hbar$ and $\hbar^2$ terms. The uncertainty relation\,\eqref{6.13}
demands that the phase plane `average' of this expression (hermitian
matrix) with $W(q,p)$ as weight function be nonnegative. After this
`averaging', the leading term is no longer a one-dimensional
projector; moreover, the pure imaginary antisymmetric part coming from
this part of $C(q,p)$ being singular, Lemma 2 of Section II does not 
apply to the real symmetric part of
$\tilde{\Omega}^{(1)}(\hat{\rho})$. In any event, \eqref{6.13} again constrains the extent to
which $W(q,p)$ can become negative. 

\section{Concluding Remarks}
In this paper we have set up a systematic procedure to obtain covariant 
uncertainty relations 
for general quantum systems. It applies equally well to continuous variable systems and  
 to systems described by finite-dimensional  Hilbert spaces, and even to 
systems based on the tensor product of the two, and consists 
of two ingredients\,: the choice of a collection of observables, 
and the action of unitary symmetry operations on them. We have shown that the uncertainty 
relations are automatically covariant---preserved in content---under every 
symmetry operation. 
 
We have applied this to two important special cases\,: the 
fluctuations and covariances in coordinates and momenta of an $n$-mode 
 canonical system; and to the set of all 
hermitian operator `monomials' in canonical variables $\hat{q},\,\hat{p}$ 
of a single mode system.  These are both generalisations 
of the Schr\"{o}dinger-Robertson UP 
in two distinct directions. The latter generalisation has been treated for definiteness 
using the Wigner distribution method.

 We hope to have set up a robust yet flexible 
formalism which can be applied  to all quantum systems, in particular
to composite, for instance bipartite, systems. In such a case, by
judicious choices of the operator sets $\{\hat{A}_a \}$ of Section II,
one can devise tests for entanglement, exhibiting covariance under
corresponding local symmetry operations. If for a bipartite system the
operator $\hat{\rho}^{\,{\rm PT}}$\,\cite{peres96}, arising from partial transpose of a
physical state $\hat{\rho}$, violates any uncertainty relation, the
presence of entanglement in $\hat{\rho}$ follows\,\cite{recent3,solomonnpt12}. 
A systematic analysis along these lines of higher order moments in the bipartite 
multi-mode case will be presented elsewhere, keeping in
mind that our general methods are applicable for both discrete and
continuous variable systems, and even to composite 
systems consisting of either or both types as subsystems.

\end{document}